\documentclass[pra,twocolumn,showpacs,preprintnumbers,amsmath,amssymb,superscriptaddress,aps,10pt]{revtex4-1}

\usepackage{graphicx}
\usepackage{dcolumn}
\usepackage{amsmath}
\usepackage{textcomp}
\usepackage{color}
\usepackage{braket}

\usepackage[caption=false]{subfig}

\begin{document}

\date{\today}


\title{Exceptionally strong correlation-driven charge migration and attosecond transient absorption spectroscopy}
\author{Maximilian Hollstein}
\affiliation{Department of Physics, University of Hamburg, Jungiusstra{\ss}e 9, D-20355 Hamburg, Germany}
\affiliation{The Hamburg Centre for Ultrafast Imaging, Luruper Chaussee 149,
22761 Hamburg, Germany}
\author{Robin Santra}
\affiliation{Center for Free-Electron Laser Science, DESY, Notkestra{\ss}e 85, 22607 Hamburg, Germany}
\affiliation{Department of Physics, University of Hamburg, Jungiusstra{\ss}e 9, D-20355 Hamburg, Germany}
\affiliation{The Hamburg Centre for Ultrafast Imaging, Luruper Chaussee 149,
22761 Hamburg, Germany}
\author{Daniela Pfannkuche}
\affiliation{Department of Physics, University of Hamburg, Jungiusstra{\ss}e 9, D-20355 Hamburg, Germany}
\affiliation{The Hamburg Centre for Ultrafast Imaging, Luruper Chaussee 149,
22761 Hamburg, Germany}

\begin{abstract}
We investigate theoretically charge migration following prompt double ionization of a polyatomic molecule (C$_2$H$_4$BrI) and find that for double ionization, correlation-driven charge migration appears to be particularly prominent, i.e., we observe exceptionally rich dynamics solely driven by the electron-electron interaction even in the situation when the electrons are emitted from outer-valence orbitals. These strongly correlated electron dynamics are witnessed in the theoretically determined time-resolved transient absorption cross section. Strikingly, features in the cross section can be traced back to electron hole populations and time-dependent partial charges and hence, can be interpreted with surprising ease. Remarkably, by taking advantage of element specific core-to-valence transitions, the hole population dynamics can be followed both in time and space. With this, not only do we report the high relevance of correlation-driven charge migration following double ionization but our findings also highlight the outstanding role of attosecond transient absorption spectroscopy (ATAS) as one of the most promising techniques for monitoring ultrafast electron dynamics in complex molecular systems.
\end{abstract}
\maketitle
The dynamics of electrons govern elementary processes such as chemical reactions and charge transport in molecules, biological systems and solid state samples. The observation and understanding of these processes is the central goal of ultrafast science. 
  With the advent of attosecond pulses, it is becoming possible to steer and to probe these electron dynamics on their natural timescale \cite{Hentschel2001,RevModPhys.81.163}. In this context, of particular interest is charge migration which represents a potentially ultrafast and efficient first step of a charge transfer which can occur even prior to the onset of a significant nuclear motion. Charge migration in molecules can take place whenever a coherent superposition of multiple electronic states is prepared. So far, this has been achieved by ionization with attosecond pulses \cite{Calegari17102014} and by strong-field ionization \cite{Krausaab2160} whereby the dynamic triggered could be well understood on the basis of a single active particle picture. Still an outstanding issue of attosecond physics is correlation-driven charge migration \cite{Cederbaum1999205,0953-4075-47-12-124002} where the electron-electron interaction is the driving force for the electron dynamics. Among a multitude of very promising techniques that have been proposed for monitoring these kind of ultrafast electron dynamics \cite{Kus2013,PhysRevA.86.053429,PhysRevLett.111.083004,PhysRevLett.111.123002,PhysRevB.91.184303,PhysRevA.88.013419}, particularly auspicious is attosecond transient absorption spectroscopy
(ATAS) \cite{Goulielmakis2010,PhysRevLett.105.143002,PhysRevLett.106.123601} combining high spectral and high temporal resolution.
ATAS has already been successfully applied to the electron dynamics in atoms \cite{Goulielmakis2010,PhysRevLett.105.143002, PhysRevA.86.063408,Chini2013,Ott2014,doi:10.1021/jp503468u} as well as to the dynamics in solid state systems \cite{Schultze2013,Schultze1348}. 
An application to correlated electron dynamics in polyatomic molecules, however,  has apparently not yet been conducted and considering that even for simple systems the interpretation of attosecond experiments is often complicated \cite{Leone2014} it is not clear to which extent ATAS can be used to gain insight into electron dynamics in systems in which electron correlation is a driving force for dynamics.

 In this Letter, we investigate charge migration that follows prompt double ionization and determine the ATA spectra for this situation. Strikingly, we observe that the electron-electron interaction already drives very diverse and complex dynamics when electrons are emitted from outer-valence orbitals. Not only do we find that these strongly correlated electron dynamics are reflected in the ATA spectra, but we also observe that features in the ATA spectra can be related to quantities such as hole populations which are comparably easy to interpret. Importantly, by taking advantage of element specific core-to-valence transitions, the associated hole population dynamics can be traced with atomic spatial resolution. Note that this feature is essential for monitoring charge transfer processes. With this, we find ATAS to be particularly suitable for the experimental evidence and investigation of correlation-driven charge migration.  
  
Subsequent to the single ionization of a molecule, electron dynamics can occur even when the ionization process is described by the prompt removal of an electron from a molecular orbital so that the state of the ionic system prepared by the ionization is given by a one-hole (1h) configuration, i.e., a state resulting from the annihilation of an electron that initially occupies a Hartree-Fock orbital from the Hartree-Fock ground state. This is due to the fact that the Coulomb interaction couples the 1h configurations to the two-hole one-particle (2h1p) configurations \cite{Cederbaum1999205}. Therefore, even in the absence of ground state correlations---as the result of final state correlation and orbital relaxation---a 1h configuration can represent a coherent population of multiple ionic eigenstates and hence, can be a non-stationary state.  This is typically the case for those 1h configurations that represent an inner-valence hole \cite{:/content/aip/journal/jcp/118/9/10.1063/1.1540618}. In contrast to single ionization, for double ionization, charge migration driven by the electron-electron interaction can already occur as the result of the mutual repulsion of the two induced electron holes. That is, when a double ionization process prepares a state that can be described by the annihilation of two electrons in Hartree-Fock orbitals from the Hartree-Fock groundstate, dynamics can take place due to the hole-hole repulsion. These dynamics driven by the hole-hole interaction can occur even for outer-valence ionization whereas, as for single ionization, processes involving additional particle-hole excitations can be expected to dominate the dynamics that follow inner-valence ionization. Here, we discuss both outer-valence double ionization where the dynamics driven by the hole-hole interaction dominate and inner-valence double ionization where  particle-hole excitations become relevant.  

In general, both strong-field ionization and ionization by attosecond pulses prepare the resulting ion only in a partially coherent and mixed state \cite{Goulielmakis2010,PhysRevLett.106.053003}. However, for simplicity, we restrict our considerations to pure electronic states.  To get insight how particle-hole excitations affect the dynamics and the ATA spectra, we describe the valence electron dynamics on two levels of theory. To describe the dynamics that are solely driven by the hole-hole interaction excluding particle-hole excitations, we expand the electronic wavefunction in terms of the two-hole (2h) configurations:\begin{equation}\label{MRCIexpansion}|\psi(t)\rangle = \sum_{i<j}\alpha_{i,j}(t)c_ic_j|\Phi_0\rangle\end{equation}Here, $c_i,c_j$ denote annihilation operators of Hartree-Fock spin orbitals and $|\Phi_0\rangle$ denotes the Hartree-Fock ground state of the neutral closed-shell molecule. In additional, more accurate calculations, we incorporate particle-hole excitations by expanding the wavefunction in terms of the 2h configurations and the three-hole one-particle (3h1p) configurations: \begin{equation}\label{MRCIexpansion}|\psi(t)\rangle = \sum_{i<j}\alpha_{i,j}(t)c_ic_j|\Phi_0\rangle + \sum_{a}\sum_{i<j<k}\alpha^a_{i,j,k}(t)c^+_ac_ic_jc_k|\Phi_0\rangle.\end{equation}In the following, we consider only initial states where electrons are annihilated from Hartree-Fock orbitals from the Hartree-Fock ground state determinant, i.e., the initial states are 2h configurations. In the absence of the electron-electron interaction, these states would be eigenstates of the dicationic Hamiltonian. For this reason, the dynamics discussed are solely driven by the electron-electron interaction. Since the many-body Schr\"odinger equation is a linear equation, the mechanisms described are still present when a superposition of 2h configurations is prepared.

 ATAS is based on the transmission of a spectrally broadband attosecond probe pulse with central photon energies suitable for the excitation of an electron from an inner-shell to the valence shell. After the transmission through the sample under investigation, the pulse is spectrally dispersed and analyzed. This allows one to monitor the valence electron dynamics with both high temporal and high spectral resolution. When applied to molecules, strong atomic localization of the inner shell orbitals and element specific inner-shell orbital energies make atomic resolution possible \cite{PhysRevA.88.013419,PhysRevA.90.023414}. In the work presented, the theoretical ATA spectra are obtained by following the procedure given in Ref. \cite{PhysRevA.83.033405}. Assuming unaligned molecules, we consider an orientation-averaged cross section. Details concerning theoretical and numerical treatment are given in the supplemental material. 
 
\begin{figure}[ht]
\includegraphics[width=0.5\textwidth]{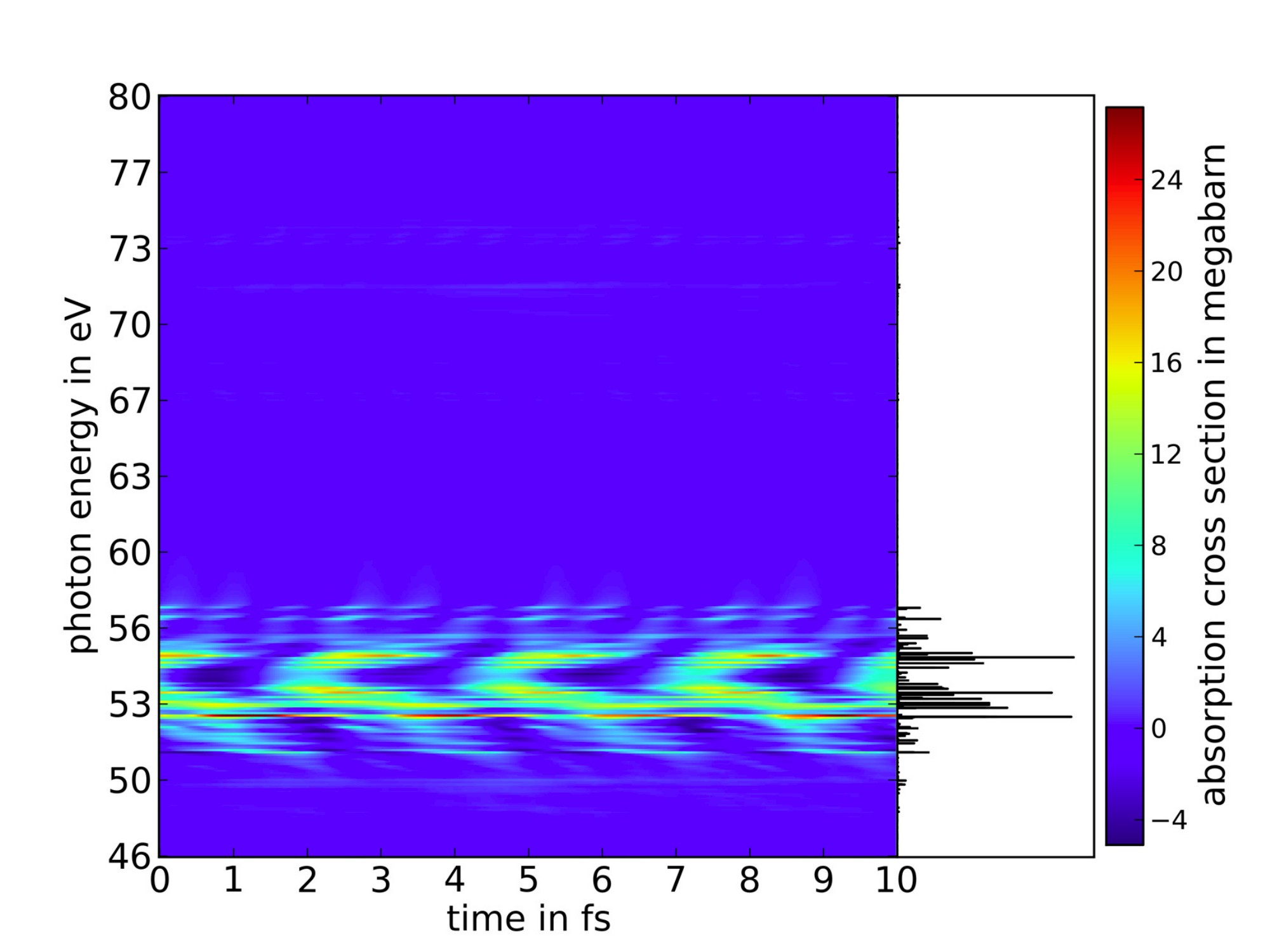}  
\caption{\label{fig:ATAS_HOMO_2h_sk}  The time-resolved absorption cross section for the situation where at $t = 0$, two electrons are suddenly removed from the HOMO. Particle-hole excitations are not taken into account. Black bars at the right side  reflect the weight of the transitions involved.}
\end{figure}

 As a specific example, we consider here the C$_2$H$_4$BrI molecule. Notably, the  iodine-4d to valence transitions and the bromine-3d to valence transitions are energetically well separated. Therefore, ATAS  allows for this molecule insight into the valence electron dynamics with atomic spatial resolution, i.e., the dynamics that occur locally at the iodine and bromine atoms, respectively, can be resolved. In the ATA spectra presented in the following, features that occur at photon energies in the region from 46 eV to 67 eV involve only final states that are associated with the promotion of an electron from an iodine-4d-type orbital to the valence shell. As in  Ref. \cite{:/content/aip/journal/jcp/141/16/10.1063/1.4898375} where a similar molecule is subject to core-to-valence spectroscopy, we refer to this spectral region as 'iodine window'. For the same reason, the region of 67 eV - 80 eV is referred to as 'bromine window' since there only transitions to final states are relevant that are associated with the promotion of an electron from the bromine-3d-type inner-shell to the valence shell.
\newline
First, we consider electronic wave packets produced via {\it outer-valence} double ionization. Specifically, the situation is considered where two electrons are suddenly removed from the HOMO. In the calculations artificially restricted to the two-hole configuration subspace, the resulting dynamics are governed by the coherent superposition of two dicationic eigenstates:\begin{equation}|\Psi_{2h}(t)\rangle = 0.74 \times e^{-iE_1t}|\Psi_1\rangle + 0.57 \times e^{-iE_2t}|\Psi_2\rangle...\end{equation} with  $E_1 - E_2 = 1.58$ eV. Both states are superpositions of mostly two spin-adapted singlet 2h configurations:\begin{equation}|\Psi_1\rangle = 0.74 \times |\text{HOMO},\text{HOMO}\rangle + 0.51 \times|\text{HOMO},\text{HOMO}-6\rangle ...\end{equation}\begin{equation}|\Psi_2\rangle = 0.57 \times |\text{HOMO},\text{HOMO}\rangle - 0.69 \times |\text{HOMO},\text{HOMO}-6\rangle ...\end{equation}
That is, the dynamics are associated with oscillatory hole populations of the HOMO and the HOMO-6. Noting that the HOMO is primarily localized at the iodine atom whereas the HOMO-6 is mostly localized at the C$_2$H$_4$ group (see supplemental material), the dynamics correspond to an oscillation of a hole between the iodine atom and C$_2$H$_4$ group with a period of $\sim2.6$ fs. This charge oscillation driven by the hole-hole interaction is reflected in the time-resolved absorption cross section in the iodine window (see Fig. \ref{fig:ATAS_HOMO_2h_sk}). 

\begin{figure}[htbp]
\includegraphics[width=0.5\textwidth]{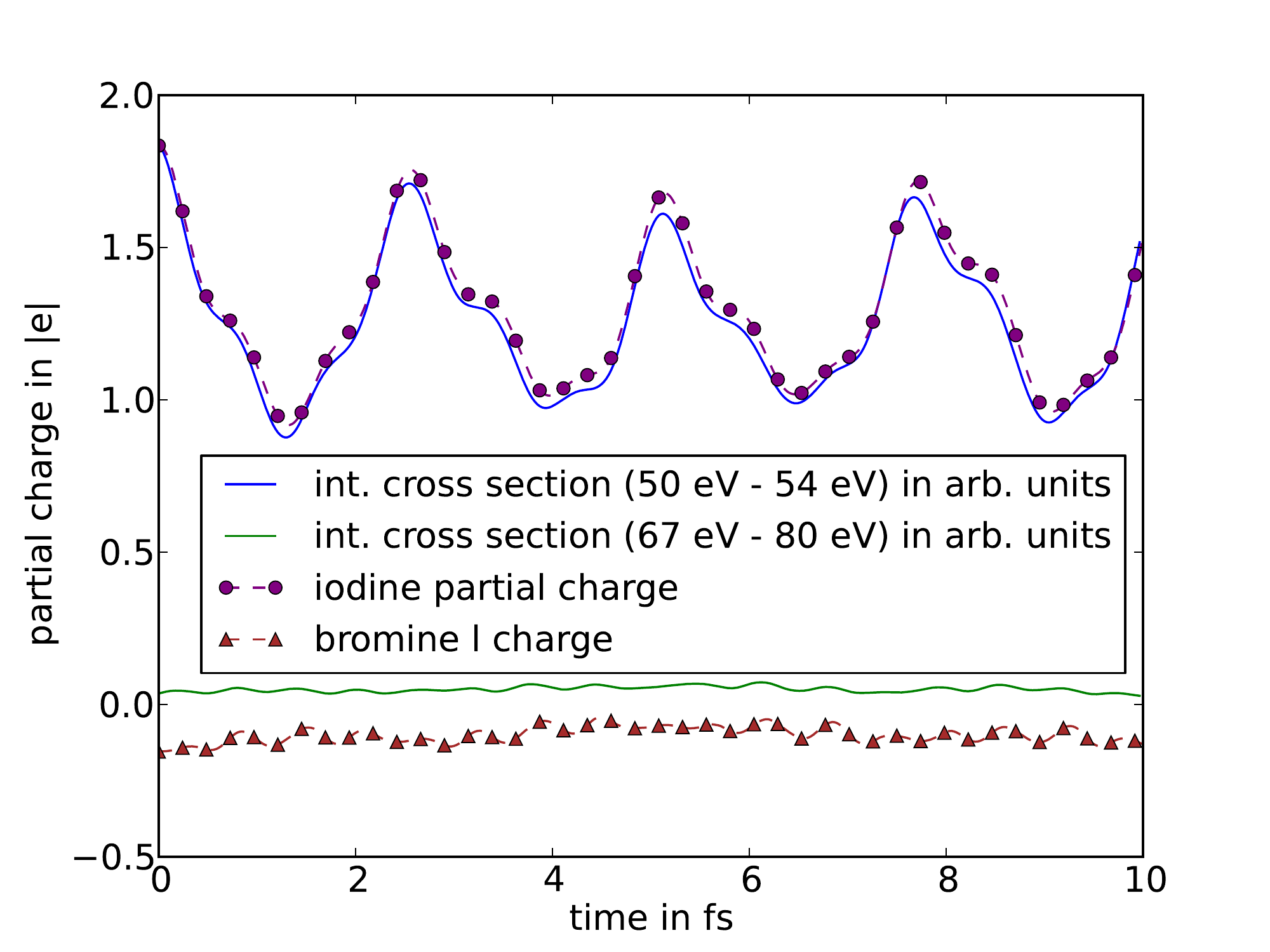}
    \caption{ The time-dependent partial charges obtained by L\"owdin population analysis in units of the elementary positive charge following the prompt removal of two electrons from the HOMO orbital, and the time-resolved absorption cross section integrated over the iodine window and the bromine window given in arbitrary units. Particle-hole excitations are not taken into account. The charge dynamics solely involving electron holes is perfectly reflected in the absorption cross section.  }
\label{fig:integrated_cross_section_and_partial_charges_2h}
\end{figure}

In fact, integration of the cross section over the iodine window ($\sim46$ eV - 67 eV) yields a quantity that is proportional to the partial charge of the iodine atom (see Fig. \ref{fig:integrated_cross_section_and_partial_charges_2h}). 
In the bromine window ($\sim67$ eV - 80 eV), however, no signal is observable, reflecting the fact that the charge dynamics are restricted to the iodine atom and the C$_2$H$_4$ group.

\begin{figure*}[!ht]
    \subfloat[\label{fig:ATAS_3h1p_HOMO}]{%
      \includegraphics[width=0.45\textwidth]{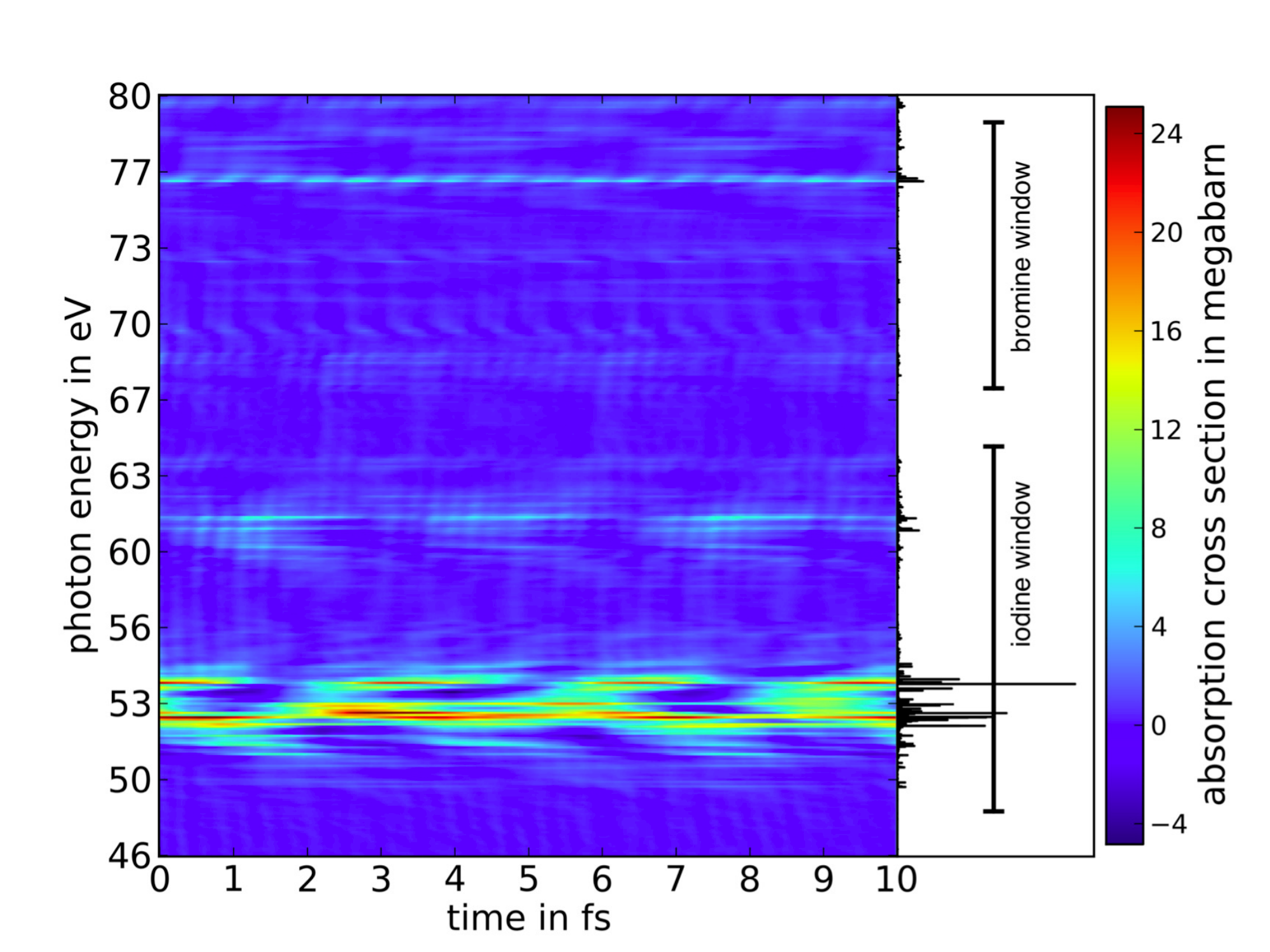}
    }
    \hfill
    \subfloat[\label{fig:ATAS_3h1p_HOMOm5}]{%
      \includegraphics[width=0.45\textwidth]{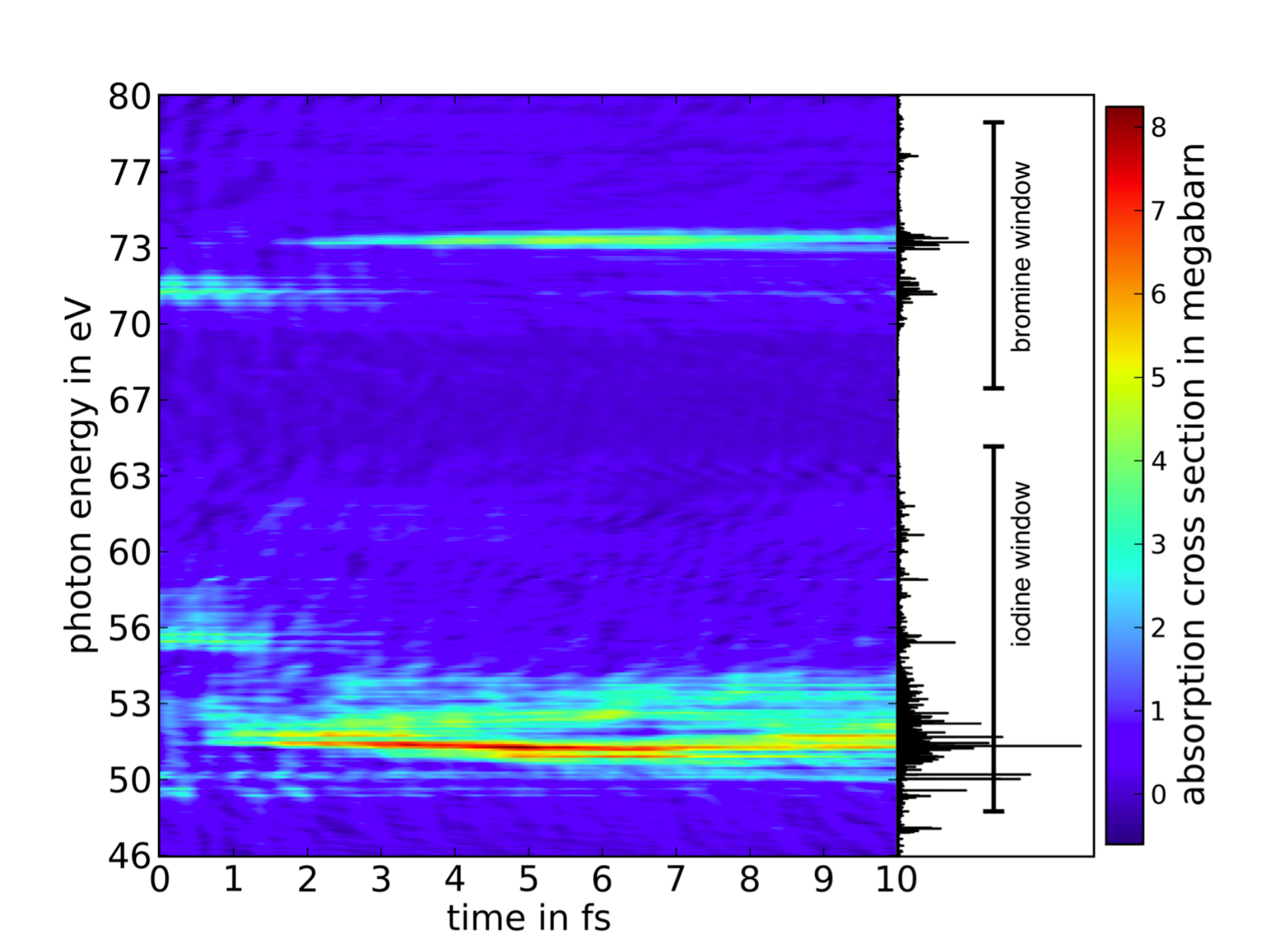}
    }
    \caption{The time-resolved absorption cross section in megabarn for the situation where a) two electrons are suddenly removed from the HOMO and b)  two electrons are suddenly removed from the HOMO-5. Particle-hole excitations are taken into account. Black bars at the right side of each panel reflect the weight of the transitions involved.}
    \label{fig:ATAS_3h1p}
  \end{figure*}

Remarkably, when including particle-hole excitations in the calculation it becomes apparent that already in this situation particle-hole excitations are relevant although the electrons are removed from the outer-valence shell. That is, a situation is considered where these kind of processes are usually assumed to be of minor importance. Here, particle-hole excitations cause prominent additional features in the time resolved absorption cross section. Besides the dominant feature at $\sim53$ eV which is already present when the calculations are restricted to the 2h subspace, at $\sim61$ eV and $\sim77$ eV new features arise (see Fig. \ref{fig:ATAS_3h1p_HOMO}). The feature at 61 eV is within the iodine window ($\sim46$ eV - 67 eV) and therefore can be related to charge dynamics at the iodine atom and is, as is the feature at $\sim53$ eV, a result of the coherent population of the two most populated states. However, in contrast to the feature at $\sim53$ eV where the final states involved have a significant 2h weight ($\geq 0.6$), for the feature at $\sim61$ eV final states are involved with a rather low 2h weight ($\leq 0.1$). 

Actually, it turns out that the feature at 61eV can be related to a charge oscillation that originates solely from the 3h1p part of the coherent population of the two most populated dicationic eigenstates. The phase of the charge oscillation described by the 3h1p part is shifted by $\pi$ with respect to the charge oscillation described by the 2h part (see supplemental material). Note that this is also reflected in the absorption cross section where the associated oscillations of the features at $\sim53$ eV and at $\sim61$ eV, respectively, exhibit the same phase shift.


The feature in the bromine window at $\sim$77 eV can be related
to a coherent superposition of the most populated state with a double ionization potential (DIP) of 27.7 eV and a group of dicationic states with dominant 3h1p character with 2h weight  smaller than 0.1 and DIPs on the order of $\sim34.5$ eV. The associated energy difference of 6.7 eV is manifested in the oscillation of the cross section with a period of $\sim610$ as. Hence, this signal at 77 eV is caused by the charge dynamics that occur locally at the bromine atom resulting from particle-hole excitations. Note that when particle-hole excitations are included in the calculations, the trends seen in the iodine and bromine partial charges are still reflected in the time-resolved cross section integrated over the iodine and the bromine window, respectively. However, the agreement is not as perfect as found for the dynamics that involve only the electron holes. As already mentioned in Ref. \cite{PhysRevA.90.023414}, this can be attributed to rather different oscillator strengths of transitions involving occupied orbitals and those transitions involving virtual orbitals. Therefore, the presence or absence of electrons in virtual orbitals is not as strongly reflected in the ATA spectra as the absence or presence of electron holes. 

When electrons are ionized from {\it inner-valence} orbitals, particle-hole excitations dominate the subsequent electron dynamics in the ionic system. 
For single ionization, this situation has been termed 'break-down of the molecular orbital picture of ionization' \cite{0022-3700-10-15-001,0022-3700-11-11-007} which is characterized by the population of many ionic eigenstates with rather low 1h weight. In the following, we consider the analogous situation for double ionization (see for instance Ref. \cite{:/content/aip/journal/jcp/92/5/10.1063/1.457894}) realized by a sudden removal of two electrons from the HOMO-5.
  The hole population dynamics (see Fig. \ref{fig:integrated_cross_section} and for more details see the supplemental material) that follow the preparation exhibit a  repopulation of the HOMO-5 by an electron, an according depopulation of the HOMO-1 and HOMO-2 on a few-fs timescale ($\sim5$ fs)  and a very fast depopulation of the HOMO-4 within only 300 as. These dynamics can be observed in the time-resolved absorption cross section (see Fig. \ref{fig:ATAS_3h1p_HOMOm5}). The HOMO-5 is delocalized throughout the whole molecule (see supplemental material) so that its depopulation can be observed at photon energies in both the iodine window at $\sim56$ eV and the bromine window at $\sim72$ eV. It turns out that the cross section integrated from 55 eV to 60 eV and from 70 eV to 72 eV, respectively, shows a similar trend as the hole population of the HOMO-5. However, as shown in Fig. \ref{fig:ATAS_3h1p}  excellent agreement is obtained in both cases rather with the sum of the hole populations of the HOMO-5 and the HOMO-4 and not solely the hole population of the HOMO-5. This can be attributed to the very similar character of these two orbitals. That is, both orbitals are completely delocalized throughout the whole molecule so that both orbitals have large overlap with both the bromine and the iodine atom. 

In contrast, the HOMO-1 and the HOMO-2 can be distinguished since the HOMO-1 is most exclusively localized at the iodine atom whereas the HOMO-2 is localized at the bromine atom. Their populations are correlated with the features in the ATA spectrum at $\sim52$ eV and at $\sim73$ eV, respectively.  The very prominent population of the HOMO-1 is proportional to the absorption cross section integrated over the very prominent feature at 50 eV to 54 eV whereas the population of the HOMO-2 is proportional to the  cross section integrated from 72 eV to 75 eV (see  Fig.
\ref{fig:integrated_cross_section}). 

\begin{figure}[!ht]
    \subfloat[\label{}]{%
      \includegraphics[width=.5\linewidth]{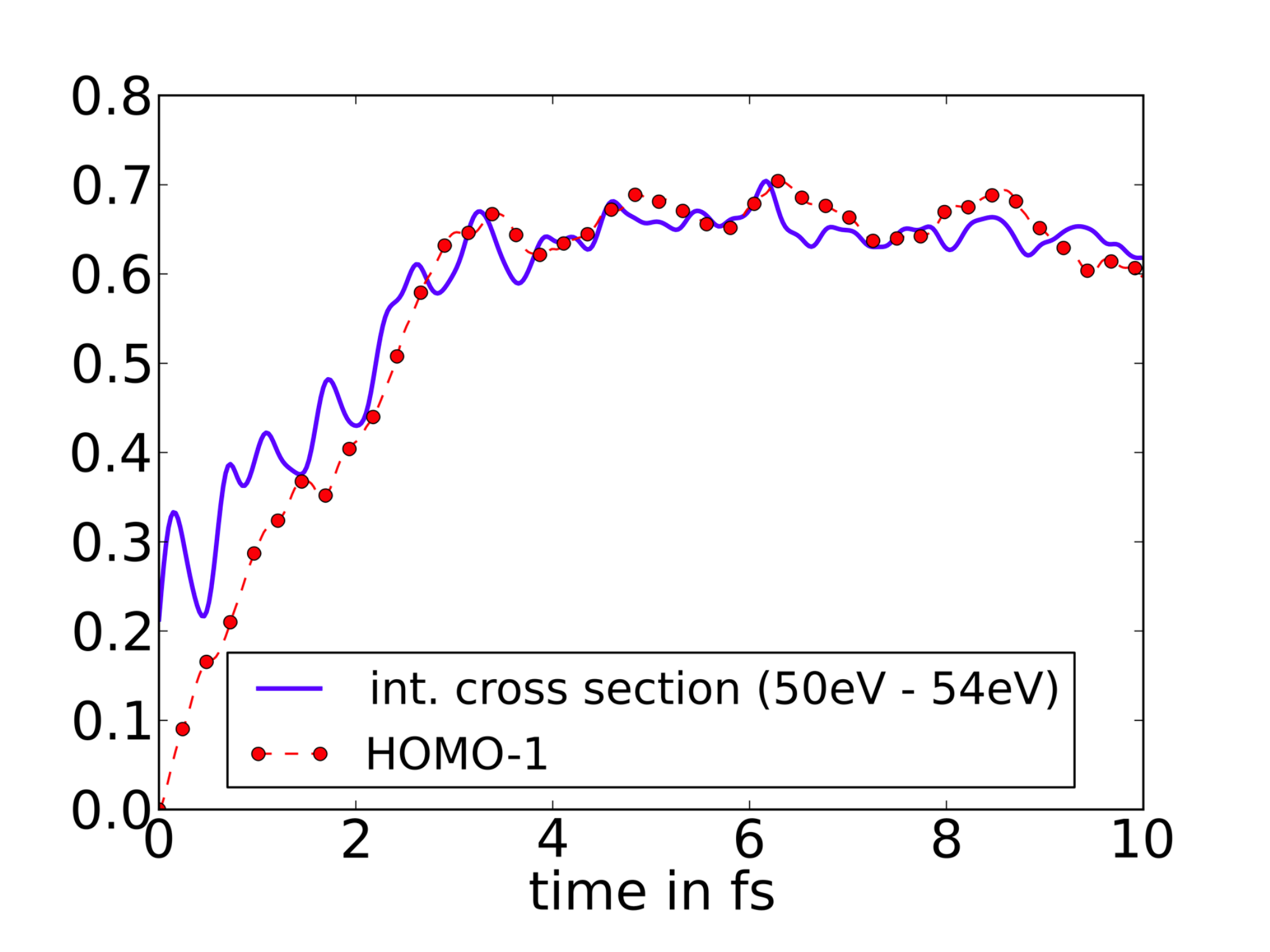}
    }
        \subfloat[\label{}]{%
      \includegraphics[width=.5\linewidth]{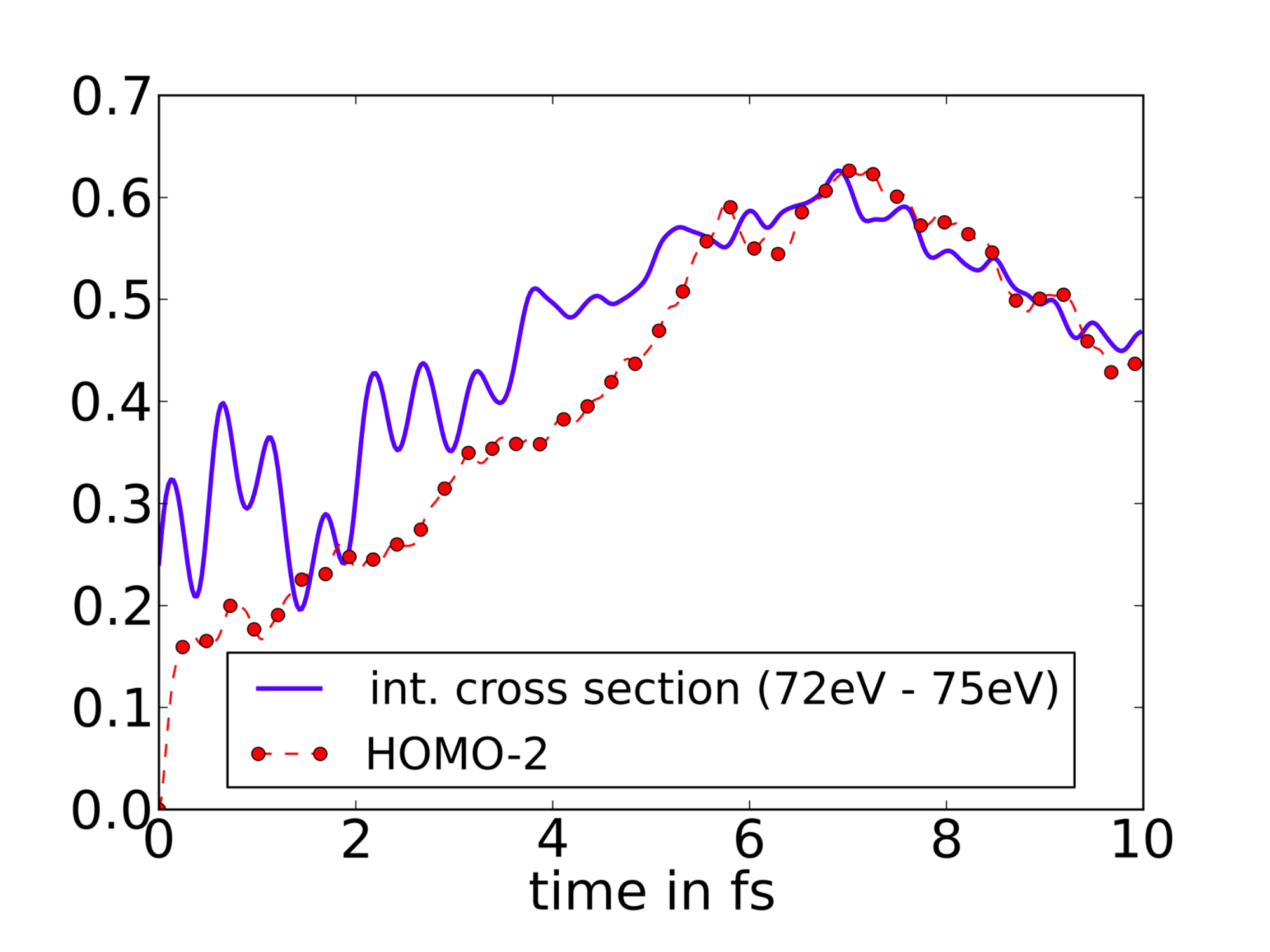}
    }\\
        \subfloat[\label{}]{%
      \includegraphics[width=.5\linewidth]{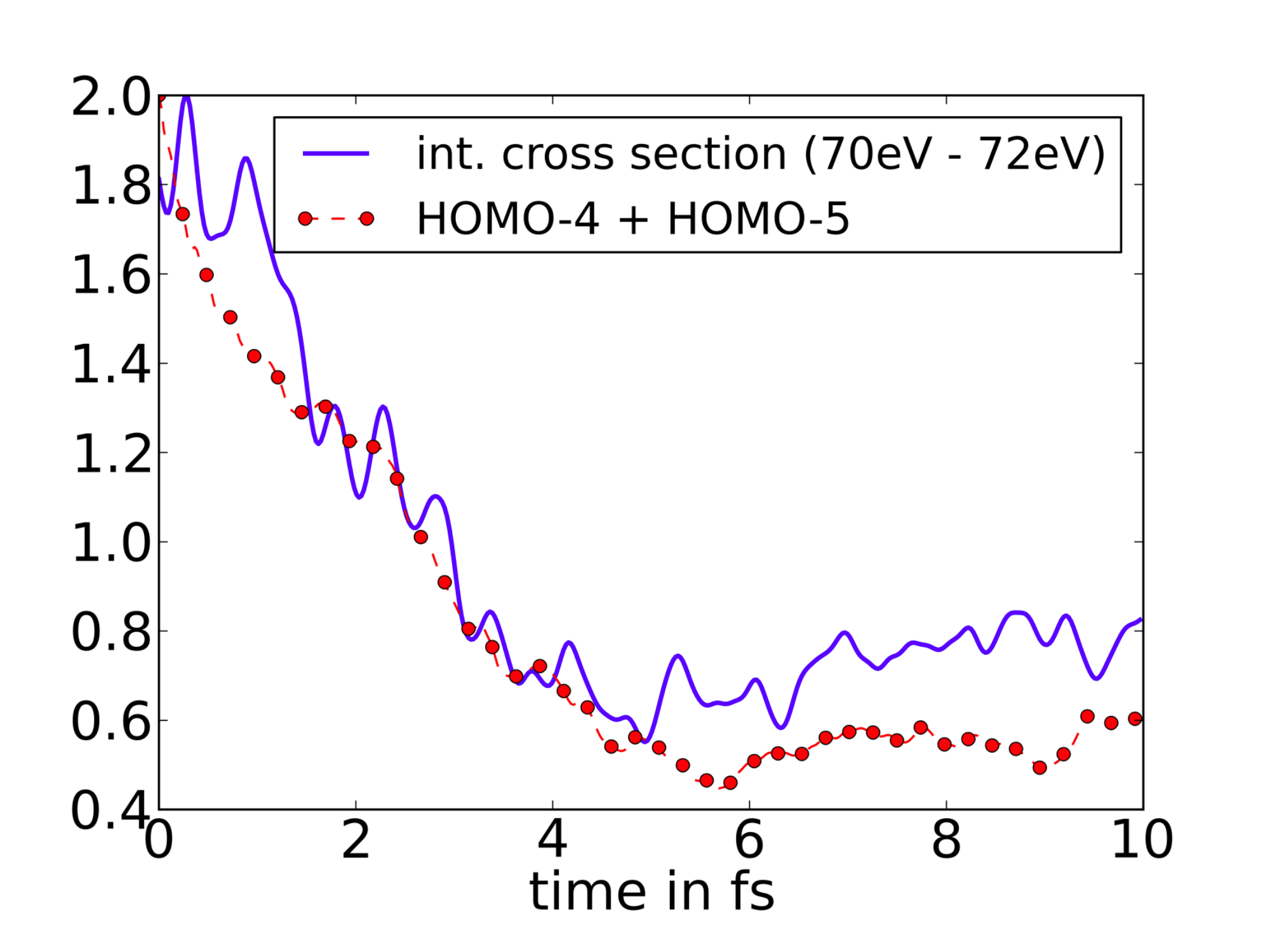}
    }
        \subfloat[\label{}]{%
      \includegraphics[width=.5\linewidth]{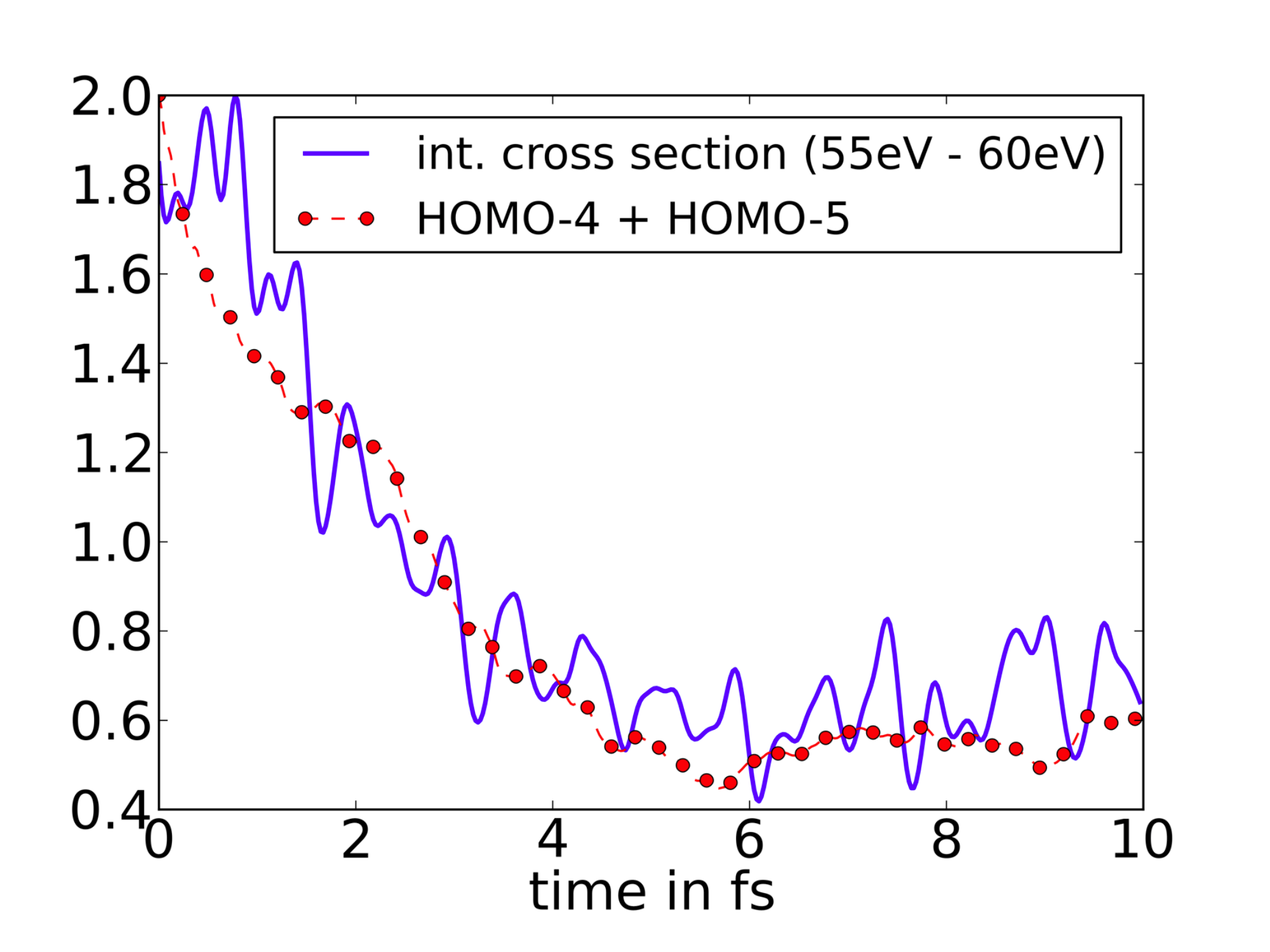}
    }
    \caption{The time-resolved absorption cross section in
arbitrary units (see Fig. \ref{fig:ATAS_3h1p}) integrated over selected spectral regions  and time-dependent hole populations for the situation where two electrons are suddenly removed from the HOMO-5. The time-dependent hole populations are reflected in the piecewise integrated absorption cross section (see text).}
\label{fig:integrated_cross_section}
\end{figure}
Remarkably, while the complicated substructure of the feature at $\sim50$ eV - 54 eV reflects the involvement of many dicationic eigenstates and rather complicated electron dynamics, the hole dynamics can rather simply be extracted from the ATA spectra. Notably, features associated with the emission of radiation  (i.e. features where the absorption cross section is negative) known from atomic ATA spectra \cite{PhysRevA.83.033405} are missing. This appears to be related to the involvement of very many transitions with only slightly different transition energies. The overlap of these many lifetime-broadened transition lines apparently causes a primarily positive cross section.

To conclude, we have demonstrated that in contrast to single ionization, for double ionization, correlation-driven charge migration is particularly prominent,  and exceptionally rich electron dynamics can already follow outer-valence double ionization. By showing that these  dynamics can be well monitored by ATAS, we have also presented the first theoretical predictions concerning ATAS applied to the correlated electron dynamics in a polyatomic molecule. Our findings suggest that even in the situation where the electron dynamics are strongly correlated, features appearing in the time-resolved absorption cross section reflect the charge flow in the molecule in time and---when it is possible to make use of the element specificity of core-to-valence transitions---also in space. Compared to other more natural approaches to resolve electron dynamics with spatial resolution such as time-resolved X-ray scattering \cite{Dixit17072012,PhysRevA.89.043409} or electron diffraction \cite{PhysRevA.88.062711}, the ATA spectra allow easy and intuitive access to to the electron dynamics. Moreover, we find that ATAS allows one to differentiate the dynamics associated with the population of different excitation classes not only when these dynamics are spatially separate but even when they affect the same atom. This finding suggests that ATAS may allow the experimental demonstration of correlation-driven charge migration.  The surprisingly straightforward interpretability of the ATA spectra and the fact that ATAS is already feasible \cite{Goulielmakis2010,Wirth2013149}, makes ATAS, in our opinion, the most promising technique for monitoring correlated electron dynamics in complex molecular systems. \section{Acknowledgments}
 Funding by the excellence cluster 'The Hamburg Centre for Ultrafast Imaging - Structure, Dynamics, and Control of Matter on the Atomic Scale' is gratefully acknowledged.

\newpage 
\bibliography{references}

\end{document}


\date{\today}

\pacs{}

\title{Exceptionally strong correlation-driven charge migration and attosecond transient absorption spectroscopy - supplemental material}
\author{Maximilian Hollstein}
\affiliation{Department of Physics, University of Hamburg, Jungiusstra{\ss}e 9, D-20355 Hamburg, Germany}
\affiliation{The Hamburg Centre for Ultrafast Imaging, Luruper Chaussee 149,
22761 Hamburg, Germany}
\author{Robin Santra}
\affiliation{Center for Free-Electron Laser Science, DESY, Notkestra{\ss}e 85, 22607 Hamburg, Germany}
\affiliation{Department of Physics, University of Hamburg, Jungiusstra{\ss}e 9, D-20355 Hamburg, Germany}
\affiliation{The Hamburg Centre for Ultrafast Imaging, Luruper Chaussee 149,
22761 Hamburg, Germany}

\author{Daniela Pfannkuche}
\affiliation{Department of Physics, University of Hamburg, Jungiusstra{\ss}e 9, D-20355 Hamburg, Germany}
\affiliation{The Hamburg Centre for Ultrafast Imaging, Luruper Chaussee 149,
22761 Hamburg, Germany}
\begin{abstract}
\end{abstract}
\maketitle

\section{Valence electron dynamics - Numerical Treatment} \label{sec:method}
Here, we provide information concerning  the computational method employed for the description of the valence electron dynamic investigated in the paper. In the situations considered, the initial state of the valence electrons of the dication is a two-hole configuration:\begin{equation}\label{eq:THconfig}|\psi(t = 0)\rangle = c_ic_j|\Phi_0\rangle \end{equation}
where $|\Phi_0\rangle$ denotes the Hartree-Fock ground state and $c_ic_j$ denote annihilation operators of Hartree-Fock spin orbitals. In the absence of the electron-electron interaction, these states would be eigenstates of the dicationic Hamiltonian so that the subsequent dynamic---if existing---is the result of the mutual interaction of the electrons captured in the dicationic molecule. Here, our  starting point for the theoretical description and the numerical treatment of the valence electron dynamic that follows the preparation is the representation of the electronic wavefunction in terms of a multi-reference configuration interaction expansion:\begin{equation}\label{MRCIexpansion}|\psi(t)\rangle = \sum_{i<j}\alpha_{i,j}(t)c_ic_j|\Phi_0\rangle + \sum_{a}\sum_{i<j<k}\alpha^a_{i,j,k}(t)c^+_ac_ic_jc_k|\Phi_0\rangle...\end{equation} An approximation of this wavefunction by an expansion that is restricted to the two-hole configurations $c_ic_j|\Phi_0\rangle$ is possible when the valence electron dynamic is governed by the dynamic of the two induced electron holes.\begin{equation}\label{FirstTwoHoleexpansion}|\psi(t)\rangle \approx \sum_{i<j}\alpha_{i,j}(t)c_ic_j|\Phi_0\rangle.\end{equation} However, electron correlation and orbital relaxation can in general result in the promotion of electrons to initially unoccupied orbitals (virtual orbitals). As indicated in equation \ref{MRCIexpansion}, excitations of electrons to virtual orbitals can be taken into account by including systematically higher excitations classes such as the 3h1p configurations $c^+_ic_jc_kc_l|\Phi_0\rangle$, the 4h2p configurations and so on until one ends up with a fully correlated description of the electron dynamic (full-CI). In the paper, we restrict our considerations to electron correlation and relaxation effects that are associated with the promotion of a single electron to the virtual orbitals.  Accordingly, we consider a time-dependent multi-reference configuration interaction singles expansion in the form of: \begin{equation}\label{Threeholeexpansion}|\psi(t)\rangle \approx \sum_{i<j}\alpha_{i,j}(t)c_ic_j|\Phi_0\rangle + \sum_{a}\sum_{i<j<k}\alpha^a_{i,j,k}(t)c^+_ac_ic_jc_k|\Phi_0\rangle.\end{equation}
With this, insight into final state correlation and relaxation can be obtained systematically by variation of the number of virtual orbitals taken into account. When no virtual orbitals are included in the calculation, an approximation in the form of equation \ref{FirstTwoHoleexpansion} is obtained whereas correlation and orbital relaxation effects associated with the promotion of a single electron to the virtual orbitals are taken into account when a finite number of virtual orbitals are included.
In order to obtain approximations of the exact wavefunction in the form of the wavefunctions given in equation \ref{Threeholeexpansion}, we propagate the initial state numerically exactly within the respective subspace. The time-evolved states are determined by: \begin{equation}|\psi(t)\rangle = e^{-iHt}|\psi(t = 0)\rangle\end{equation} where H denotes the Hamiltonian matrix restricted to the respective subspace. When the initial state is represented in terms of the eigenvectors $|I\rangle$ of H as denoted in equation \ref{eq:ini_state_in_eigenstates}\begin{equation}\label{eq:ini_state_in_eigenstates}
|\psi(t = 0)\rangle = \sum_{I}\alpha_I |I\rangle,
\end{equation}
the time-evolved state can be obtained by:\begin{equation}
|\psi(t)\rangle = \sum_{I}\alpha_Ie^{-iE_It}|I\rangle
\end{equation} Here, the coefficients $\alpha_{I}$ are the expansion coefficients of the initial state expanded in dicationic eigenstates  $|I\rangle$ and $E_I$ are the corresponding energies. In the situations considered, the subspace dimensions are on the order of 200 to 10000 so that dicationic eigenstates can be obtained by full diagonalization of the Hamiltonian matrix using LAPACK \cite{laug}. The Hartree-Fock orbitals, the one-particle integrals of the kinetic energy and, the core potentials, which are needed for the calculation of the many-body Hamiltonian matrix elements, are taken from the quantum chemistry software MOLCAS \cite{Karlstrom2003222}; the Coulomb matrix elements are obtained from the LIBINT library \cite{libint}. For the iodine atom and the carbon atom, we employed effective core potentials and associated basis sets from \cite{Peterson2006} (iodine), \cite{Bergner1993}(carbon) and respectively \cite{:/content/aip/journal/jcp/119/21/10.1063/1.1622924}(bromine) whereas for the hydrogen atoms 6-31G basis sets were employed. The molecular geometries are determined by MOLCAS on the Hartree-Fock SCF wave function level. \section{Time-resolved attosecond transient absorption cross section} \label{sec:ATAS}

For a numerical determination of the time-resolved absorption cross section, we applied the theory of attosecond transient absorption spectroscopy developed in Ref. \cite{PhysRevA.83.033405}. According to Ref. \cite{PhysRevA.83.033405}, for molecules that are oriented, the attosecond transient absorption cross section can be obtained in the short-pulse approximation and by using Beer's law  by:
\begin{equation}\label{eq::crossZ}\begin{split}\sigma_{Z}(\omega,t) = \frac{4\pi\omega}{c}\text{Im} \sum_{I,I'}\alpha^*_{I}e^{iE_It}\alpha_{I'}e^{-iE_{I'}t}\\\times\sum_{F}\langle I|\hat{Z}|F\rangle\langle F|\hat{Z}|I'\rangle\\\times\Big(\frac{1}{ \tilde{E}_F - E_I - \omega} + \frac{1}{\tilde{E}^*_F - E_I'  + \omega}\Big)\end{split}
\end{equation} Here, $\hat{Z}$ is the component of the electric dipole operator along the laser polarization. \newline In the situation where the molecules are completely unaligned (this situation is assumed in the paper), the cross section can be obtained by averaging over all possible orientations with respect to the laser polarization: \begin{equation}\label{averagecs}\begin{split}\sigma(\omega,t) = \frac{\omega}{c}\text{Im} \int_0^{\pi}\int_0^{2\pi} d\phi d\theta \text{sin}(\theta)\\\sum_{I,I'}\alpha^*_{I}e^{iE_It}\alpha_{I'}e^{-iE_{I'}t}\\\times\sum_{F}\langle I|\hat{e}_r(\theta,\phi)\vec{R}|F\rangle\langle F|\hat{e}_r(\theta,\phi)\vec{R}|I'\rangle\\\times\Big(\frac{1}{ \tilde{E}_F - E_I - \omega} + \frac{1}{\tilde{E}^*_F - E_I'  + \omega}\Big)\\\\= \frac{\sigma_{X}(\omega,t) + \sigma_{Y}(\omega,t) + \sigma_{Z}(\omega,t)}{3}\end{split}
\end{equation} Here, $\hat{e}_r(\theta,\phi)$ is the radial unit vector and $\vec{R}$ is the electric dipol operator.
$\sigma_{X},\sigma_{Y},\sigma_{Z}$ are the cross sections, according to equation \ref{eq::crossZ}, obtained for three orthogonal polarization directions. 

The inner-shell vacancies decay within a few femtoseconds (see for instance Ref. \cite{:/content/aip/journal/jcp/97/11/10.1063/1.463468}). For this reason, we assign the final states a finite life time $\frac{1}{\Gamma}$ which enters equation \ref{eq::crossZ} in form of complex final state energies $\tilde{E}_{F} = E_{F} - \frac{i\Gamma}{2}$. Here, we assumed a decay width of the inner-shell vacancies on the order of the atomic one i.e. $\Gamma \sim 100\text{meV}$. For the discussion in the paper, this appears to be reasonable whereas for a direct comparison with experimental data a more accurate values might be employed. To include final states where a single electron is promoted from the iodine 4d inner-shell and from the bromine 3d inner-shell to the valence shell, we constructed the subspaces as described in the previous section. For a reduction of computation time, however, we constructed the subspaces under the restriction that at maximum only a single hole is in an inner-shell orbital. 
 Since in the situations considered, states with a double ionization potential (DIP) larger than 50 eV are insignificantly populated  (i.e. their populations are on the order of 0.001 and smaller), for a significant reduction in computation time, we restricted the sum in equation \ref{eq::crossZ} to those states with a DIP lower than 50 eV. We also restrict the sum over the final states $|F\rangle$ to those states that are relevant in the spectral region considered (that is, for photon frequencies ranging from $\omega_{min} = 46 $eV to $\omega_{max} = 80 $eV), i.e. those states $|F\rangle$ for which the condition  $ \omega_{min} < |\tilde{E}_{F} - E_I|<\omega_{max} $ is fulfilled for all populated dicationic eigenstates $|I\rangle$. Furthermore, final states $| F \rangle$ for which the transition weights  $|\alpha_I|^2 |\langle I|Z|F\rangle|^2$ for all populated dicationic eigenstates $| I \rangle $ fall below $10^{-8} $(in atomic units) are found to be irrelevant for the spectra determined and are consequently excluded from the calculation.

\begin{figure*}[!ht]
    \subfloat[\label{}]{%
      \includegraphics[width=.5\linewidth]{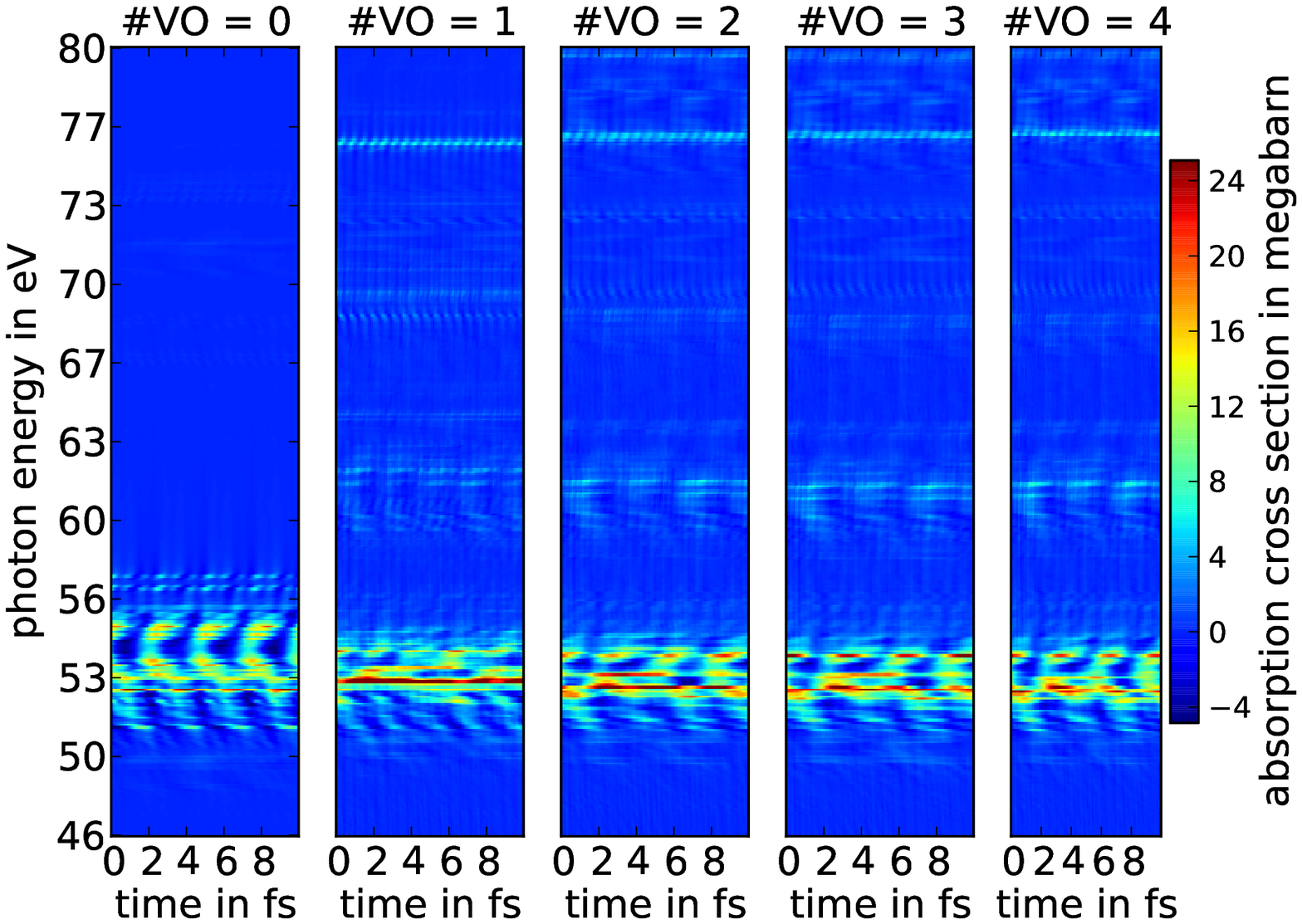}
    }
    \subfloat[\label{}]{%
      \includegraphics[width=.5\linewidth]{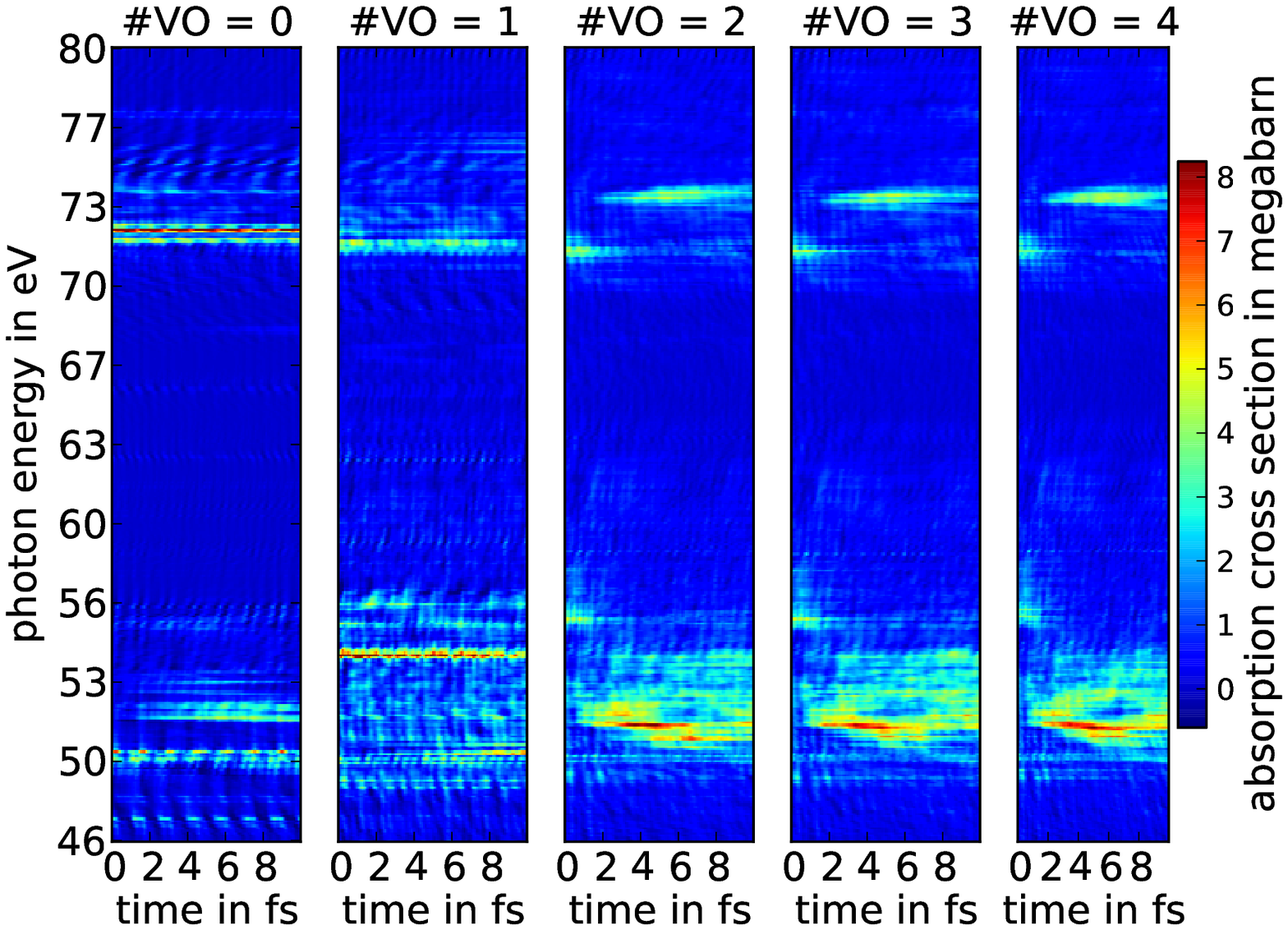}
    }
\caption{The time-resolved absorption cross section in dependence of the number of virtual orbitals (VO) that are included in the calculation (see section \ref{sec:method}). Here, the situations are considered where  a) two electrons are removed from the HOMO and b) two electrons are removed from the HOMO-5. Already by inclusion of two virtual orbitals, the features discussed arise. In the paper, we showed the time-resolved absorption cross section obtained from calculations in which 4 virtual orbitals were included.}
\label{fig:molecular_orbitals_HOMO}
\end{figure*}

In the paper, we showed the time-resolved absorption cross section in which 4 virtual orbitals were included in the calculation. However, the features caused by particle-hole excitations discussed in the paper appear already when only two virtual orbitals are taken into account - the increase from 2 to 4 virtual orbitals reveals small changes (see Fig. \ref{fig:molecular_orbitals_HOMO} ) but no new features arise.

%
\section{Valence electron motion} 
\begin{figure}[ht]

\subfloat[\label{fig:spectral_decomp_HOMO}]{%
      \includegraphics[width=.5\linewidth]{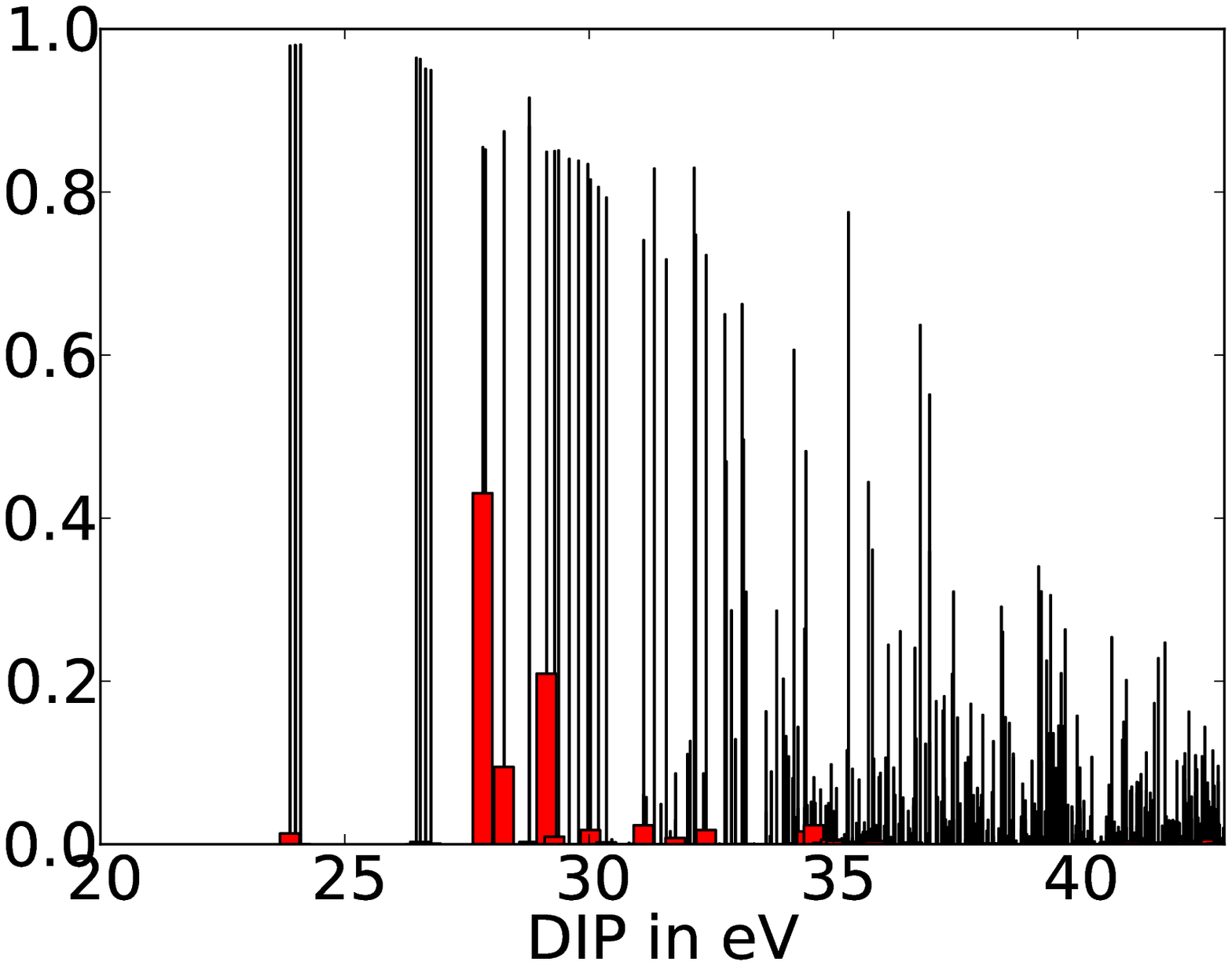}
    }
    \subfloat[\label{fig:spectral_decomp_HOMOm5}]{%
      \includegraphics[width=.5\linewidth]{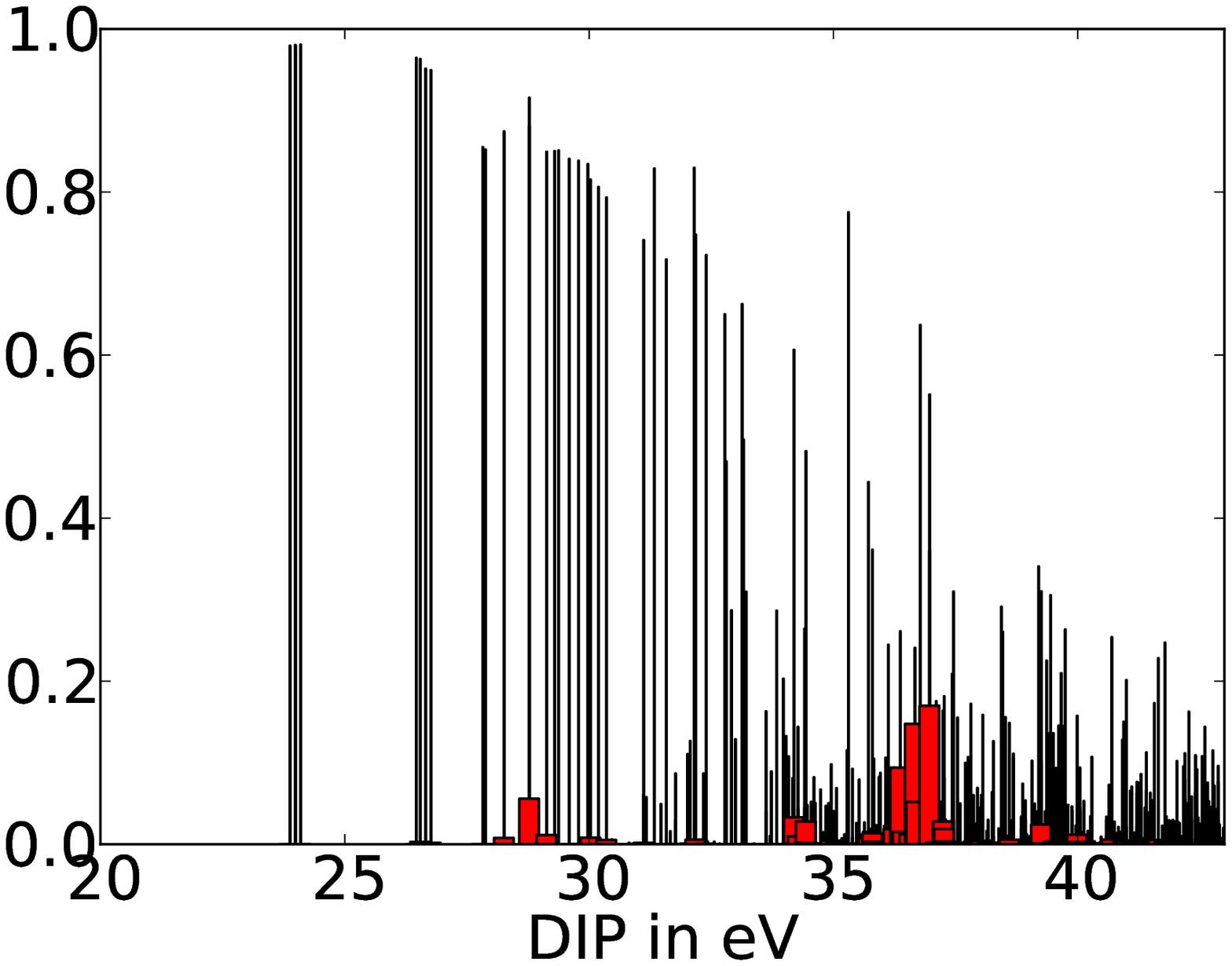}
    }

\caption{The spectral decomposition of the initial states (red bars) and the two-hole weights (black bars) associated with the corresponding dicationic eigenstates. a) The 2h configuration representing two holes in the HOMO. b) The 2h configuration representing two holes in the HOMO-5.}\label{fig:spectral_decomp}
\end{figure}

Here, we provide additional information concerning the electron dynamics discussed in the paper. The spectral decomposition of the initial states considered are shown in Fig. \ref{fig:spectral_decomp}.
\begin{figure}[ht]

      \includegraphics[width=1.0\linewidth]{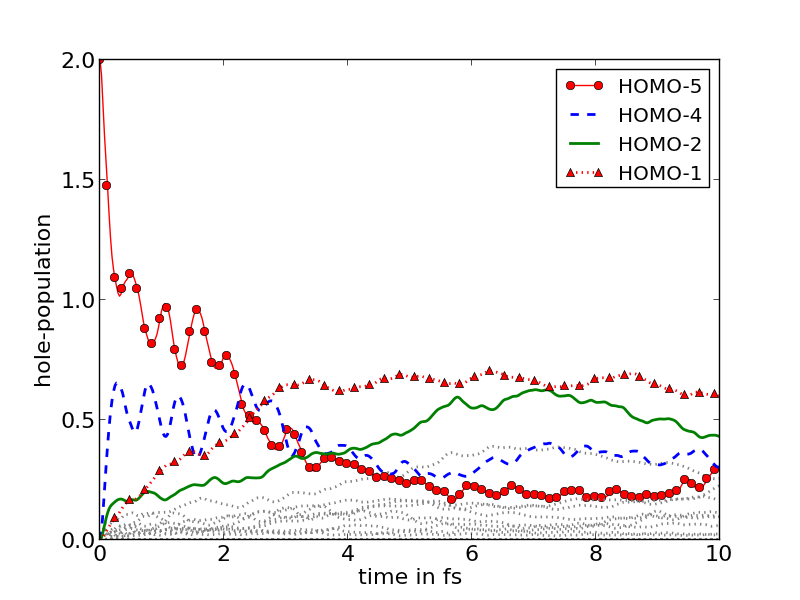}

\caption{The time-dependent hole populations following the preparation of the 2h configuration that is obtained by the annihilation of two electrons that initially occupy the HOMO-5 orbital from the Hartree-Fock ground state.}\label{fig:hole_populations}
\end{figure}
There, one can observe that the state resulting from the annihilation of the two electrons that initially occupy the HOMO from the Hartree-Fock ground state, is primarily a superposition of two dicationic eigenstates (see Fig. \ref{fig:spectral_decomp_HOMO}).  In contrast, when the two electrons that initially occupy the HOMO-5 are annihilated, many dicationic eigenstates are populated that have a comparably low 2h weight (see Fig. \ref{fig:spectral_decomp_HOMOm5}). The time-dependent hole populations that follow the preparation of this state are shown in Fig. \ref{fig:hole_populations} exhibiting a fast repopulation of the HOMO-5.

In both cases, primarily dicationic states are populated that have energies below the triple ionization threshold which is estimated on the level of Koopmans theorem at a DIP of $\sim43$ eV.

\begin{figure*}
\subfloat[\label{fig:difference_charge_coherence_8_13_3h1p_part}]{%
      \includegraphics[width=.5\linewidth]{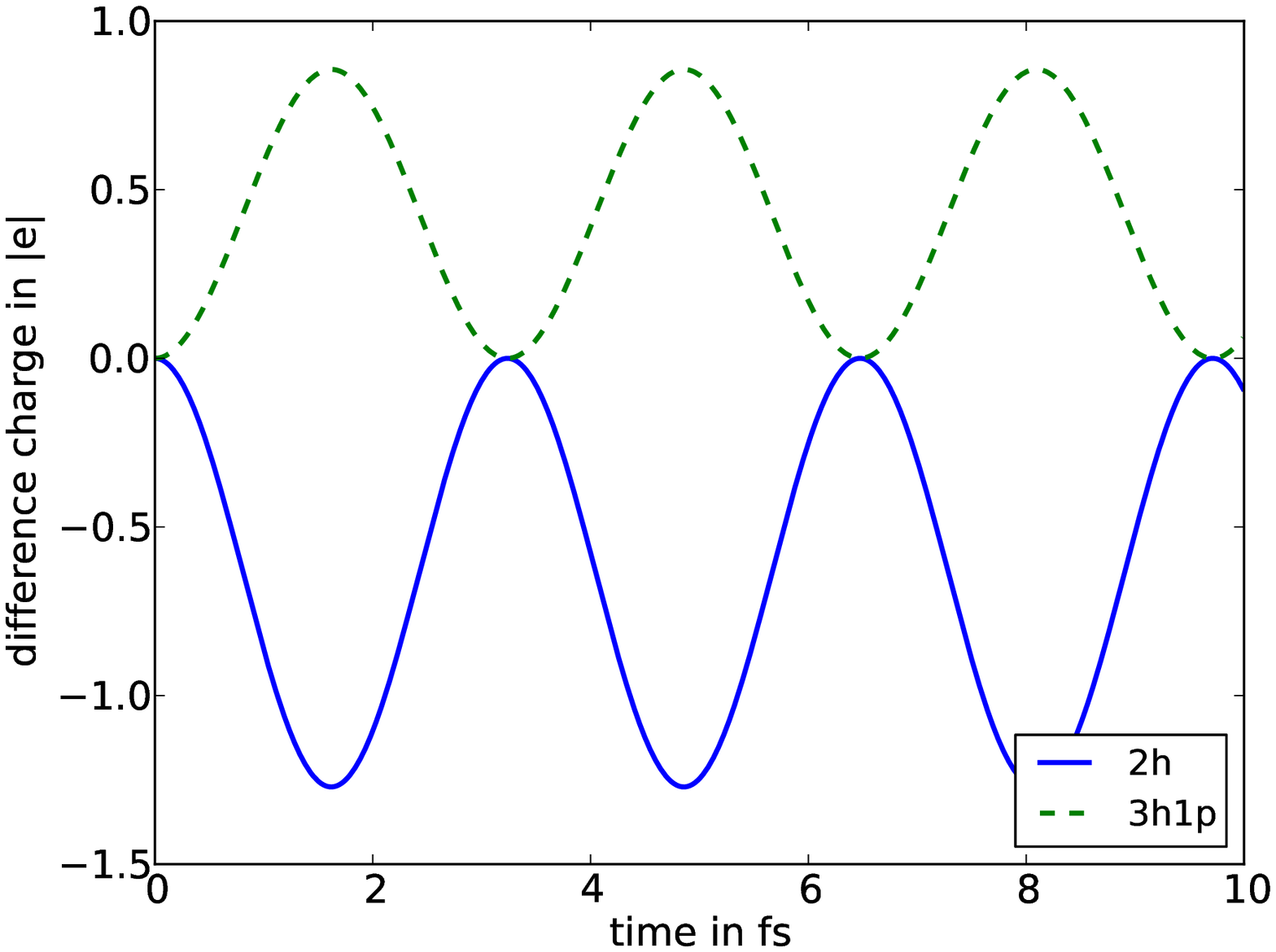}
    }
        \subfloat[\label{fig:ATAS_HOMO_3h1p_5VO_sk_sgcoherence}]{%
      \includegraphics[width=.5\linewidth]{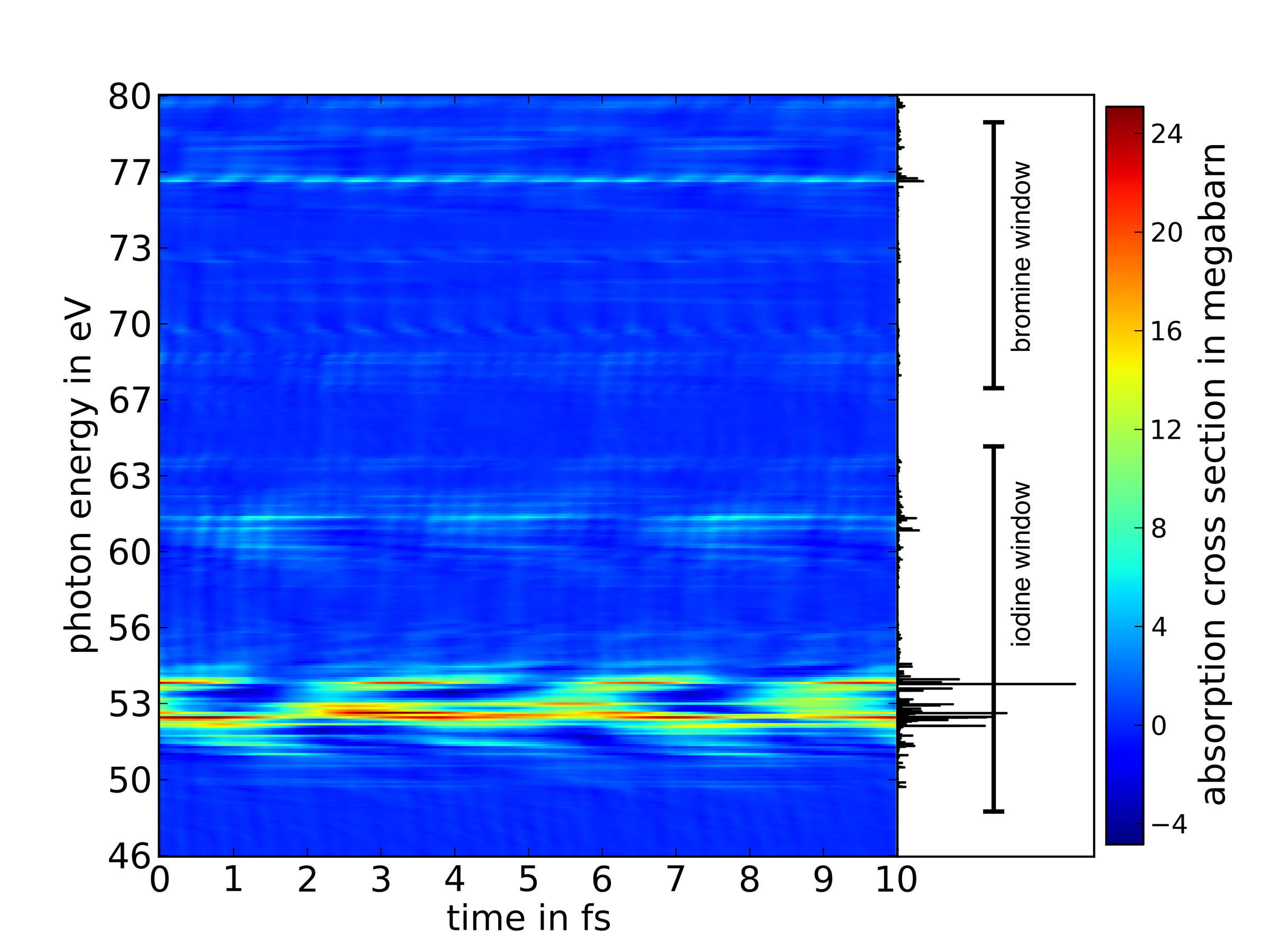}
    }
    \caption{\label{fig:HOMO_shakeup_dynamics}a) The time-dependent difference partial charge (see equation \ref{eq:DQ}) obtained from the 3h1p-part and, respectively, the 2h part of coherent superposition of the two dicationic eigenstates that are mostly populated after the sudden removal of two electrons from the HOMO (see Fig. \ref{fig:spectral_decomp_HOMO}).  The charge oscillations that can be related to the population of 3h1p configurations  are reflected in the time-resolved absorption cross section shown in b) at photon energies in the vicinty of 61 eV whereas the charge oscillations related to the 2h configurations are indicated at 53 eV.}
  
\end{figure*}

\begin{figure}[ht]

    \subfloat[\label{fig:F_HOMOm1}HOMO-1]{%
      \includegraphics[width=.25\textwidth]{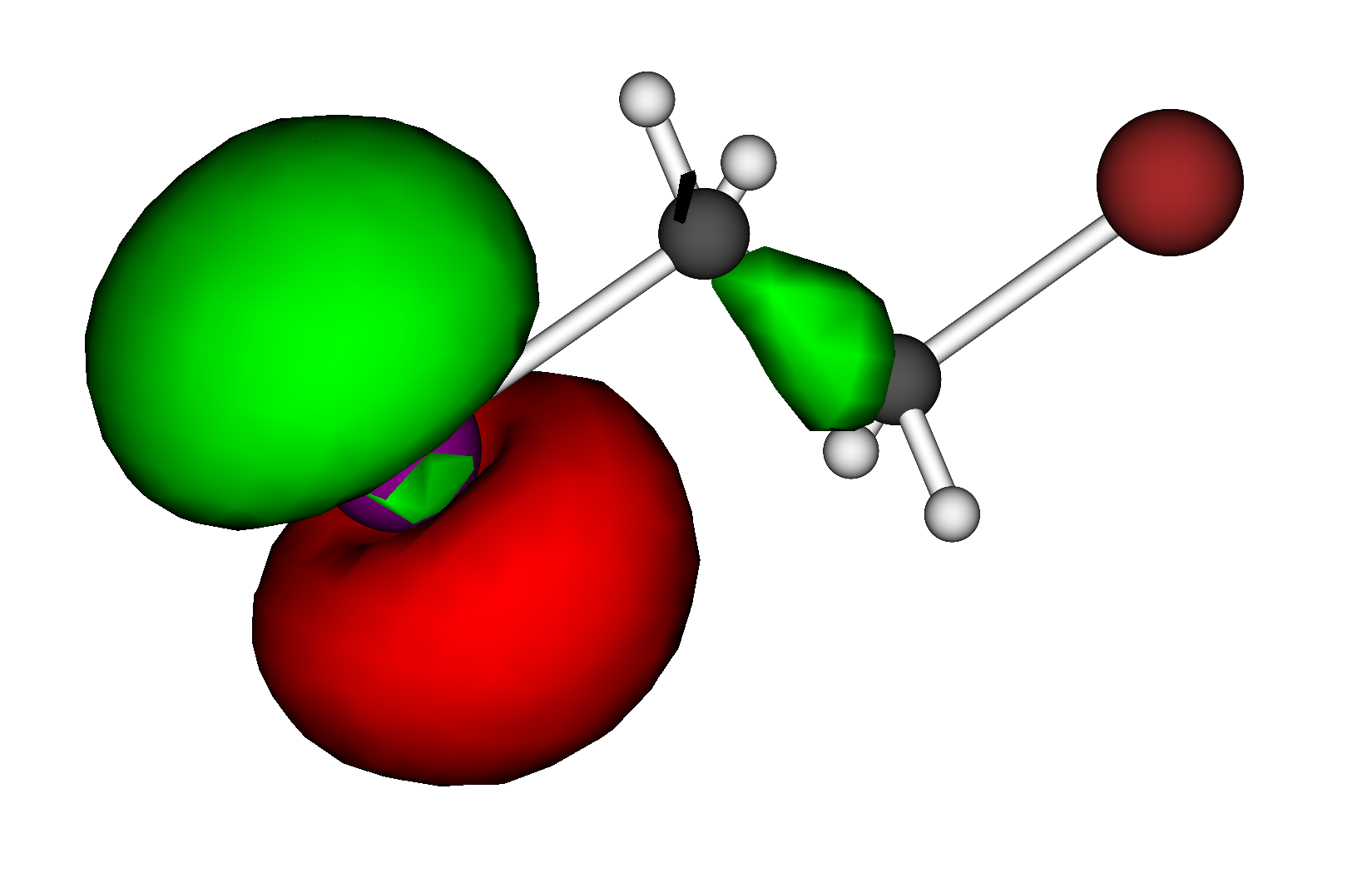}
    }
    \subfloat[\label{fig:F_HOMOm2}HOMO-2]{%
      \includegraphics[width=.25\textwidth]{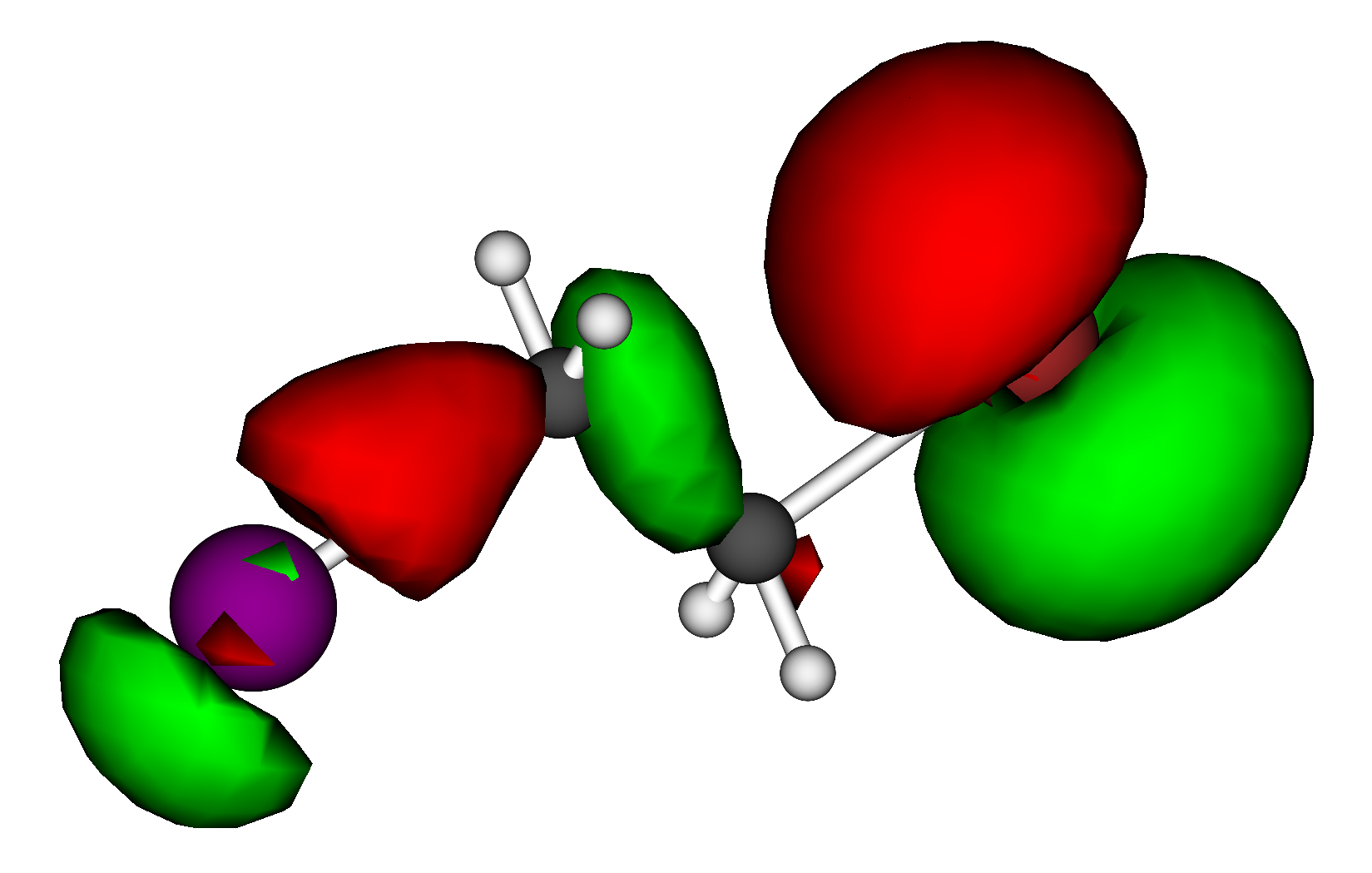}
    }
    \subfloat[\label{fig:F_HOMOm4}HOMO-4]{%
      \includegraphics[width=.25\textwidth]{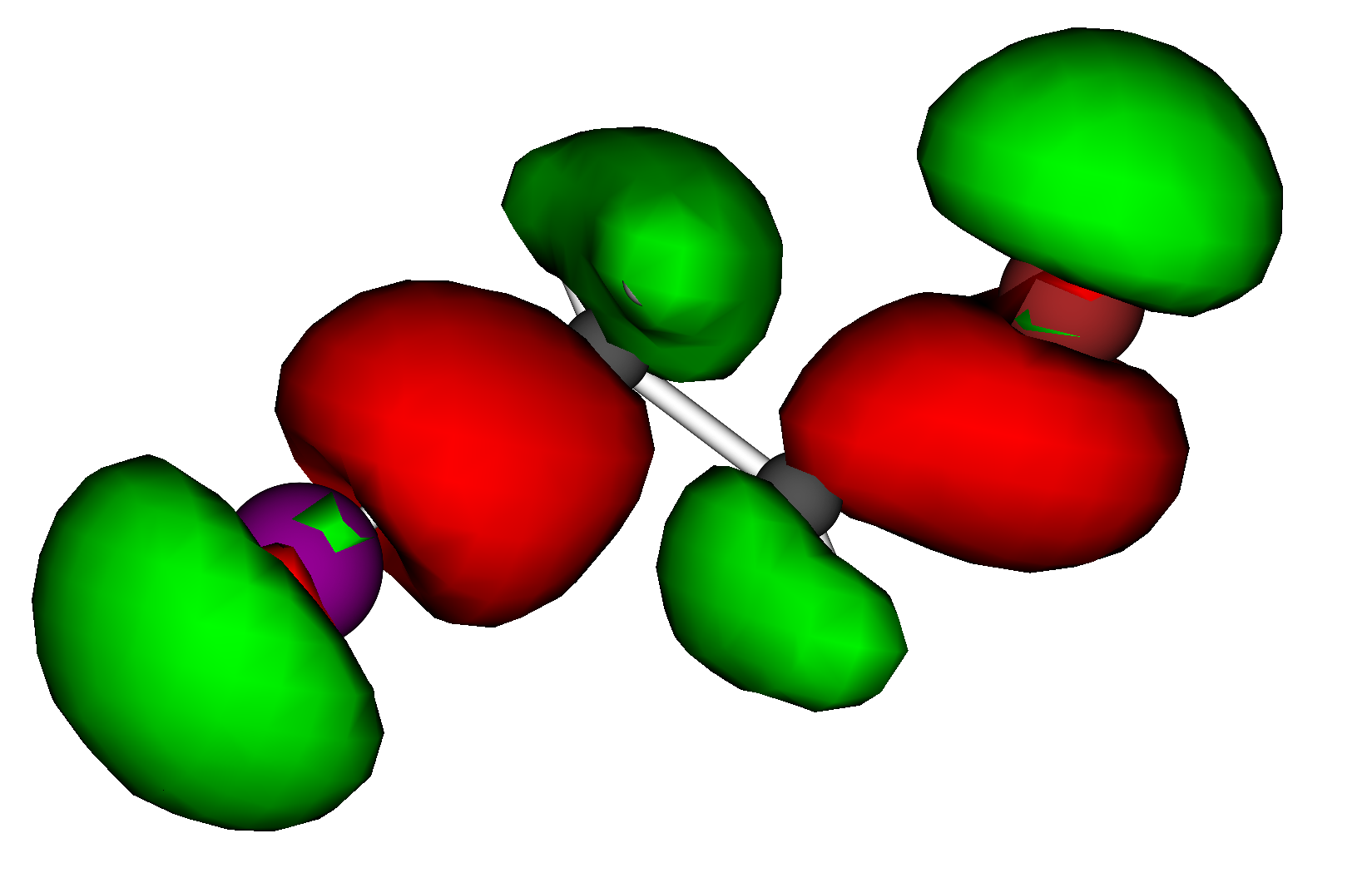}
    }
    \subfloat[\label{fig:F_HOMOm5}HOMO-5]{%
      \includegraphics[width=.25\textwidth]{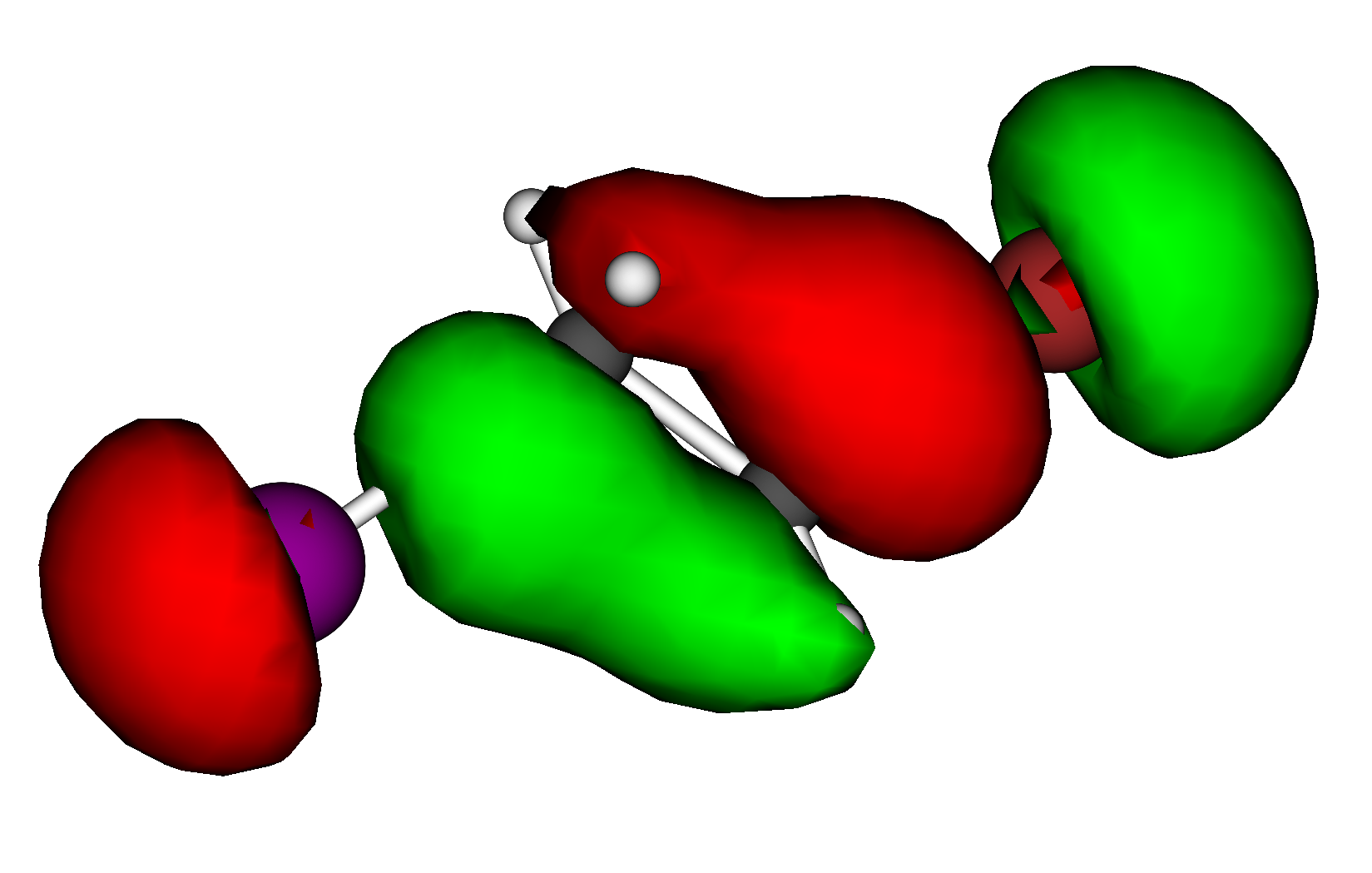}
    }\\
\subfloat[\label{fig:F_HOMO}HOMO]{%
      \includegraphics[width=.25\textwidth]{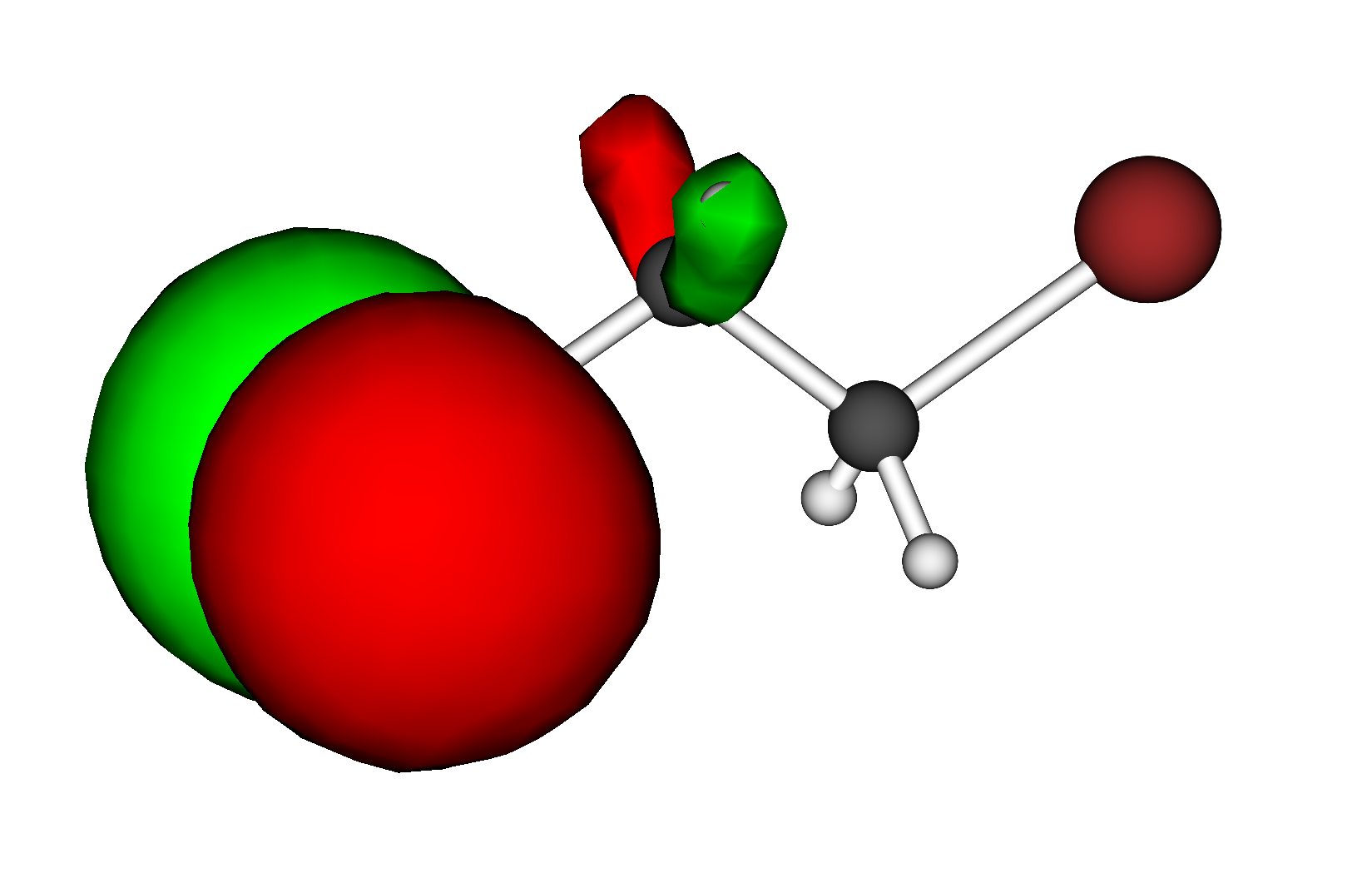}
    }
    \subfloat[\label{fig:F_HOMOm6}HOMO-6]{%
      \includegraphics[width=.25\textwidth]{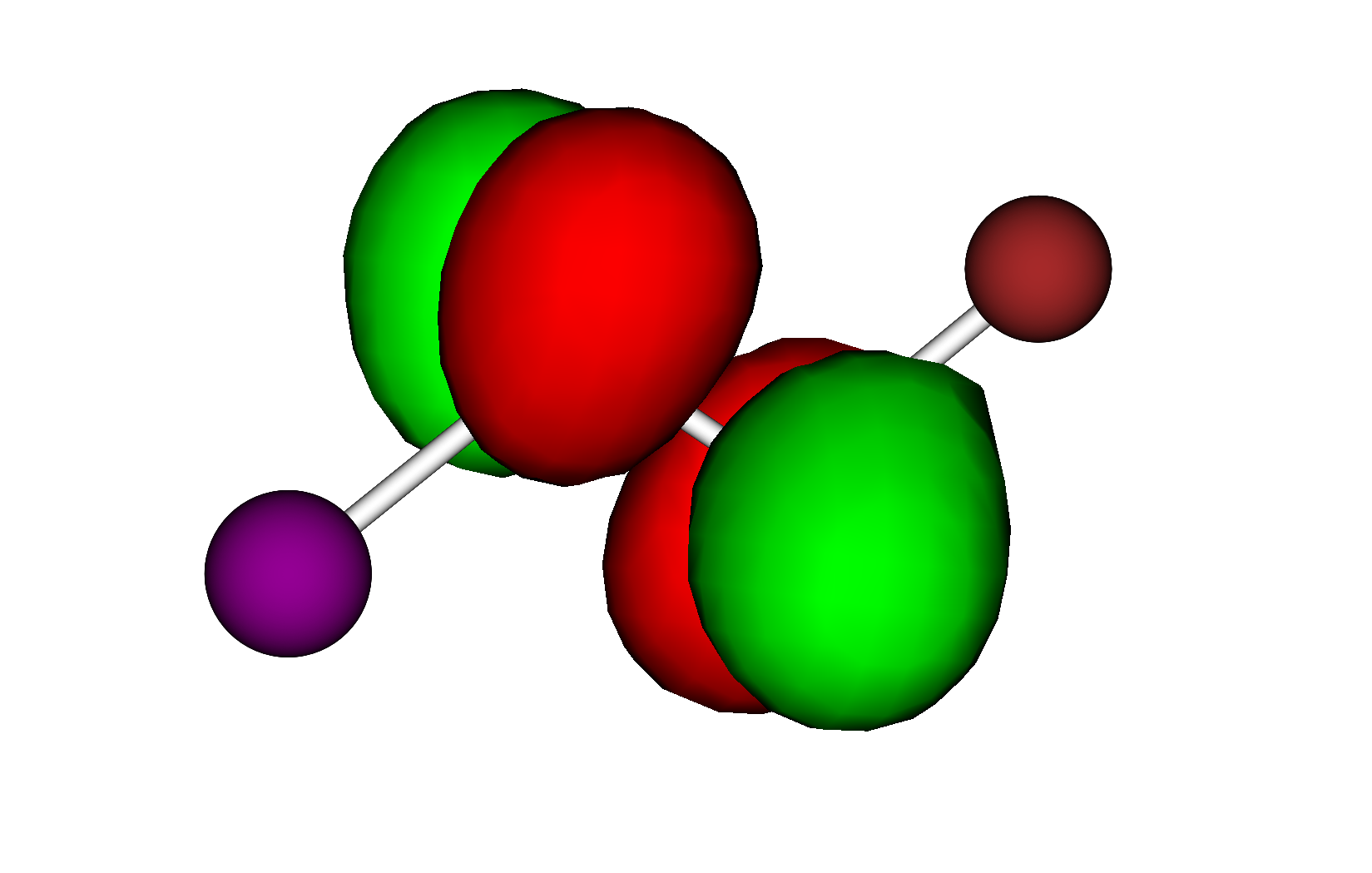}
    }

\caption{The isosurfaces of the molecular orbitals most prominently involved  in the valence electron dynamics considered in the paper.}
\label{fig:molecular_orbitals_HOMOm5}
\end{figure}

In the following, we provide additional information concerning the dynamics that follow the prompt removal of two electrons from the HOMO. As mentioned in the paper, the HOMO of C$_2$H$_4$BrI is primarily localized at the iodine atom. This is shown in Fig.  \ref{fig:F_HOMO} where one can observe that the HOMO has large overlap with an iodine p-like orbital. As discussed in the paper, the hole-hole repulsion causes oscillating hole populations of the HOMO and the HOMO-6. As can seen in Fig. \ref{fig:F_HOMOm6}), the HOMO-6 is primarily localized at the C$_2$H$_4$ group. Therefore, the hole-hole interaction drives a charge oscillation between the iodine atom and the C$_2$H$_4$ group. As mentioned in the paper, particle-hole excitations cause additional charge dynamics which are reflected in the absorption cross section (see Fig. \ref{fig:HOMO_shakeup_dynamics}). Fig. \ref{fig:difference_charge_coherence_8_13_3h1p_part} shows the time-dependent iodine difference charges
\begin{equation}\label{eq:DQ}\Delta Q(t) = Q(t) - Q(0)\end{equation} obtained separately from the 2h part and the 3h1p part of the coherent superposition of the two most populated dicationic eigenstates. Here, Q denotes the iodine partial charge obtained by  L\"owdin population analysis.  The charge oscillations that can be attributed in this way to the 2h part and the 3h1p part exhibit a phase shift of $\pi$.
Apparently, these two different charge dynamics can be observed in the time-resolved absorption cross section shown in Fig. \ref{fig:ATAS_HOMO_3h1p_5VO_sk_sgcoherence} in the iodine window (46 eV - 67 eV). 
That is, the feature at 60  eV - 63 eV and that in the vicinity of 53 eV in the absorption cross section exhibit the same time-dependence as the charge oscillations shown in Fig. \ref{fig:difference_charge_coherence_8_13_3h1p_part} - in particular, both exhibit a phase shift of $\pi$.
The interpretation that here, the dynamics associated with the population of the two distinct excitation classes are witnessed at distinct photon energies is supported by the fact that the final states relevant for the feature between 60  eV - 63 eV have only a low 2h weight whereas the final states relevant for the feature in the vicinity of 53 eV have a comparably large 2h weight.

\clearpage\bibliography{references_sm}